\documentclass[%
 aip,
 amsmath,amssymb,
 reprint,%
]{revtex4-1}

\usepackage{graphicx}
\usepackage{dcolumn}
\usepackage{bm}
\usepackage{amsmath}
\usepackage[utf8]{inputenc}
\usepackage[T1]{fontenc}
\usepackage{mathptmx}
\usepackage{etoolbox}
\usepackage{tikz}
\usepackage{siunitx}
\makeatletter
\def\@email#1#2{%
 \endgroup
 \patchcmd{\titleblock@produce}
  {\frontmatter@RRAPformat}
  {\frontmatter@RRAPformat{\produce@RRAP{*#1\href{mailto:#2}{#2}}}\frontmatter@RRAPformat}
  {}{}
}%
\makeatother
\begin{document}

\preprint{AIP/123-QED}

\title[]{Suppressing stimulated Raman side-scattering by vector light}
\author{Xiaobao Jia}
\altaffiliation{Department of Plasma Physics and Fusion Engineering and CAS Key Laboratory of
Geospace Environment, University of Science and Technology of China, Hefei, Anhui
230026, China}
\author{Qing Jia}%
 \email{qjia@ustc.edu.cn}
\affiliation{ 
Department of Plasma Physics and Fusion Engineering and CAS Key Laboratory of
Geospace Environment, University of Science and Technology of China, Hefei, Anhui
230026, China
}%

\author{Rui Yan}
\affiliation{Department of Modern Mechanics, University of Science and Technology of China, Hefei 230026, China
}%
\affiliation{Collaborative Innovation Center of IFSA, Shanghai Jiao Tong University, Shanghai
200240, China}
\author{Jian Zheng}
\email{jzheng@ustc.edu.cn}
\affiliation{ 
Department of Plasma Physics and Fusion Engineering and CAS Key Laboratory of
Geospace Environment, University of Science and Technology of China, Hefei, Anhui
230026, China
}%
\affiliation{Collaborative Innovation Center of IFSA, Shanghai Jiao Tong University, Shanghai
200240, China}


\begin{abstract}
Recently, the observations of stimulated Raman side-scattering (SRSS) in different laser inertial confinement fusion ignition schemes poses an underlying risk of SRSS on ignition. In this paper, we propose a method to use the non-uniform polarization nature of vector light to suppress SRSS and give an additional threshold condition determined by the parameter of vector light. For SRSS at 90 degrees, where the scattered electromagnetic wave travels perpendicular to the density profile, the polarization variation of the pump will change the wave vector of scattered light, thereby reducing the growth length and preventing the scattered electromagnetic wave from growing. This suppressive scheme is verified through three-dimensional particle-in-cell simulations. Our illustrative simulation results demonstrate that for linearly polarized Gaussian light, the SRSS signal occurs in the 90-degree direction fiercely. At the same time, for the vector light, there is few SRSS signal even if the condition dramatically exceeds the threshold. Furthermore, we discuss the impact of vector light on stimulated Raman and Brillouin backscattering, and two-plasma decay.
\end{abstract}

\maketitle

%

\section{\label{sec1}Introduction}
Laser Inertial Confinement Fusion (ICF) exhibits rich parametric instabilities, such as stimulated Raman scattering (SRS)\cite{forslundTheoryStimulatedScattering1975,liuRamanBrillouinScattering1974}, stimulated Brillouin scattering (SBS)\cite{forslundTheoryStimulatedScattering1975,liuRamanBrillouinScattering1974}, two-plasmon decay (TPD)\cite{RN295}, crossed-beam energy transfer (CBET)\cite{kruerEnergyTransferCrossing1996,mckinstrieTwoDimensionalAnalysis1996}, along with other secondary instabilities. SRS, whereby the incident electromagnetic wave scattered by the electron plasma wave (EPW), has been the focus of persistent attention for several decades because it reduces laser energy coupling efficiency and preheats the target capsule with hot electrons. Based on the direction of the wave vector of scattered light, SRS can be categorized as forward scattering, backward scattering, and side scattering.

SRS exhibits absolute growth (temporal amplification of initial seed) in uniform plasma while experiencing convective growth (spatiotemporal amplification of initial seed) in non-uniform plasma except for at the density of $0.25n_c$, where $n_c$ is the critical density.
In non-uniform plasma, the stimulated Raman side-scattering (SRSS) is thought to be absolutely growing and is of great significance since the scattered light tangential to the density gradient is scarcely affected by the inhomogeneity of plasma. However, despite conditions far beyond the threshold for triggering the absolute mode, little SRSS evidence was observed in 20 Th-century experiments. Mostrom\cite{mostromRamanSideScatterInstability1979} clarified the discrepancy between theory and experiment, pointing out that SRSS undergoes transverse convective growth and can only enter the absolute growth stage after the convective mode saturates by refraction from the resonance zones, and the finite width of the laser beam raises the threshold for detecting the absolute mode of SRSS\cite{mostromRamanSideScatterInstability1979,xiaoStimulatedRamanSidescattering2018}. After the 1980s, SRSS receives little attention in both experimental and theoretical investigations.

Recently, SRSS has regained attention as the experimental results demonstrated the crucial role SRSS played in indirect-drive \cite{michelMultibeamStimulatedRaman2015,Dewald2016}, direct-drive \cite{DDreview2015} as well as shock ignition \cite{PhysRevLett.98.155001} ICF ignition schemes. The collective SRSS via shared EPW was evidenced as a mechanism of hot electron generation in the indirect-drive experiment at National Ignition Facility (NIF)\cite{michelMultibeamStimulatedRaman2015,Dewald2016} and the collective SRSS via shared scattered light was observed in a direct-drive experiment performed on Omega facility\cite{PhysRevLett.117.235002}. The direct-drive experiments with a planar target in NIF showed that SRSS was the main contributor of hot electron generation in ignition-scale condition\cite{PhysRevLett.120.055001, michelTheoryMeasurementsConvective2019a}. Also, the importance of SRSS was verified in shock ignition experiments at low densities\cite{cristoforetti_2019}. Very recently, the dominance of SRSS over stimulated Raman back-scattering (SRBS) was identified at SG-II UP facility\cite{glizeMeasurementStimulatedRaman2023} in double cone ignition\cite{zhangj2020} experiments. To control the hot electrons generated by SRSS, it is necessary to mitigate the SRSS. 

In this paper, we propose suppressing SRSS by vector light \cite{VLreview2018,zhanCylindricalVectorBeams2009} featured with transversely varying polarization. The vector light is typically generated in two ways: by outputting from a designed or modified laser resonator or by using a spatial light modulator to manipulate the amplitude, phase, or both of the two orthogonally linearly polarized lights (or two left- and right-handed circularly polarized lights)\cite{wangGenerationArbitraryVector2007,wangNewTypeVector2010}. 
The suppression mechanism of polarization distribution in the cross-section of the pump beam on SRSS includes two aspects. On the one hand, at the initial stage, since the maximum growth happens where the polarization of the seed aligns with the pump, the direction of the scattered wave vector will change with the polarization variation of the pump. As a result, the scattered seeds excited at the different transverse locations are incoherent due to different polarizations, which will slow down the growth. On the other hand, the polarization variation will cease the convective growing process. The scattered light convects with group velocity in the direction of the wave vector, and the change of the directions of the wave vector and group velocity reduces the convective length so that the convective gain decreases. As is pointed out by Mostrom\cite{mostromRamanSideScatterInstability1979}, limited to the initial amplitude of the seed, the SRSS exhibits convective growth first and then absolute growth only if the convective gain is large enough. For vector light, the inhibited convective growth will prevent the scattered light from entering the absolute growth stage. Consequently, the SRSS can be strongly suppressed.

In contrast to existing mitigation schemes such as broadband light\cite{follettThresholdsAbsoluteInstabilities2019} or multicolor light\cite{zhaoMitigationMultibeamStimulated2021}, sunlight-like laser\cite{weng},  polarization smoothing\cite{PS1} or smoothing by spectral dispersion\cite{ssd}, and polarization rotation\cite{Ido}, which suppress LPIs based on the frequency spectrum, phase distribution, light intensity or temporally changing uniform polarization. In this work, we propose a scheme to inhibit LPIs using the non-uniform polarization nature of the vector light and give the parameter design of vector light to suppress SRSS, to the best of our knowledge, for the first time. As will be shown by three-dimensional PIC simulations, the SRSS does not occur even when the condition far exceeds the threshold. The non-uniform polarization nature of vector light provides additional avenues for exploring laser-plasma interaction. The remainder of this paper is structured as follows:  in Sec II, we present the PIC simulation verification of the suppressing efficacy of vector light on SRSS. The additional threshold determined by the characteristic length of vector light is derived in Sec III. Then, we discuss the impact of vector light on the backscattering and TPD, followed by a summary.

\section{\label{sec2}Simulations}
The verification of the suppressive effect on SRSS is performed by the three-dimensional PIC code EPOCH\cite{Arber_2015}. In this section, we give one of the methods of construction of vector light in PIC code and then present the contrastive simulation cases of vector light and linearly polarized Gaussian light to illustrate the suppressive effect on SRSS of vector light.

\subsection{Construction of vector light in PIC code}
\begin{figure}[htpb]   
\includegraphics[width=0.95\linewidth]{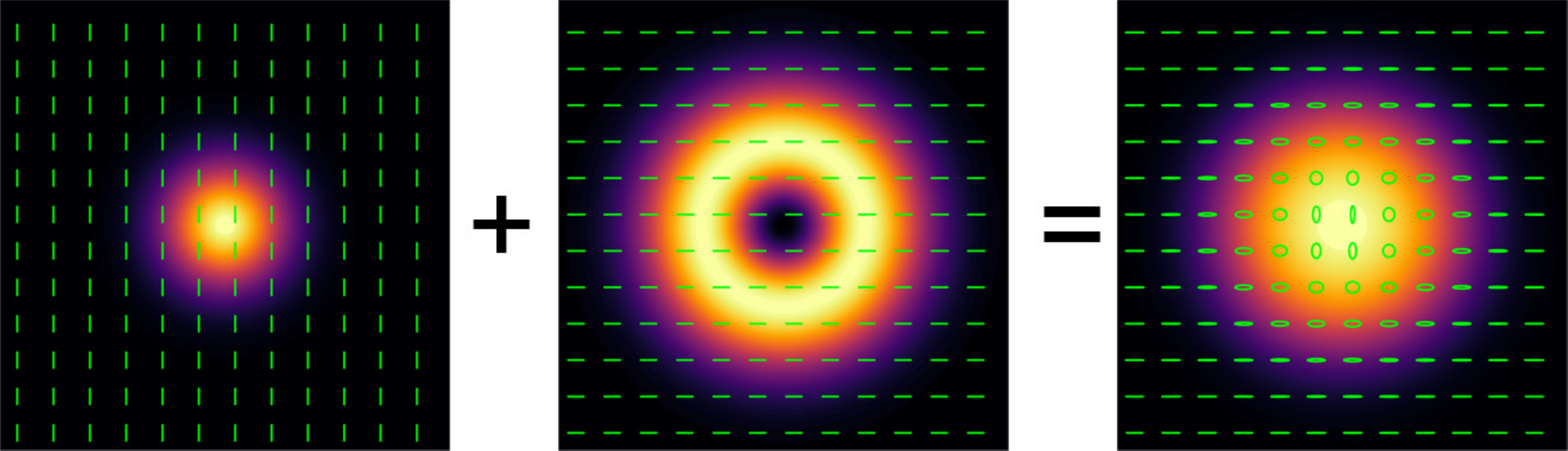}
\caption{\label{fig1} Schematic of the nonuniform polarization of vector light which has a Gaussian intensity profile.}
\end{figure}
\begin{figure}[htpb]   
\includegraphics[width=0.85\linewidth]{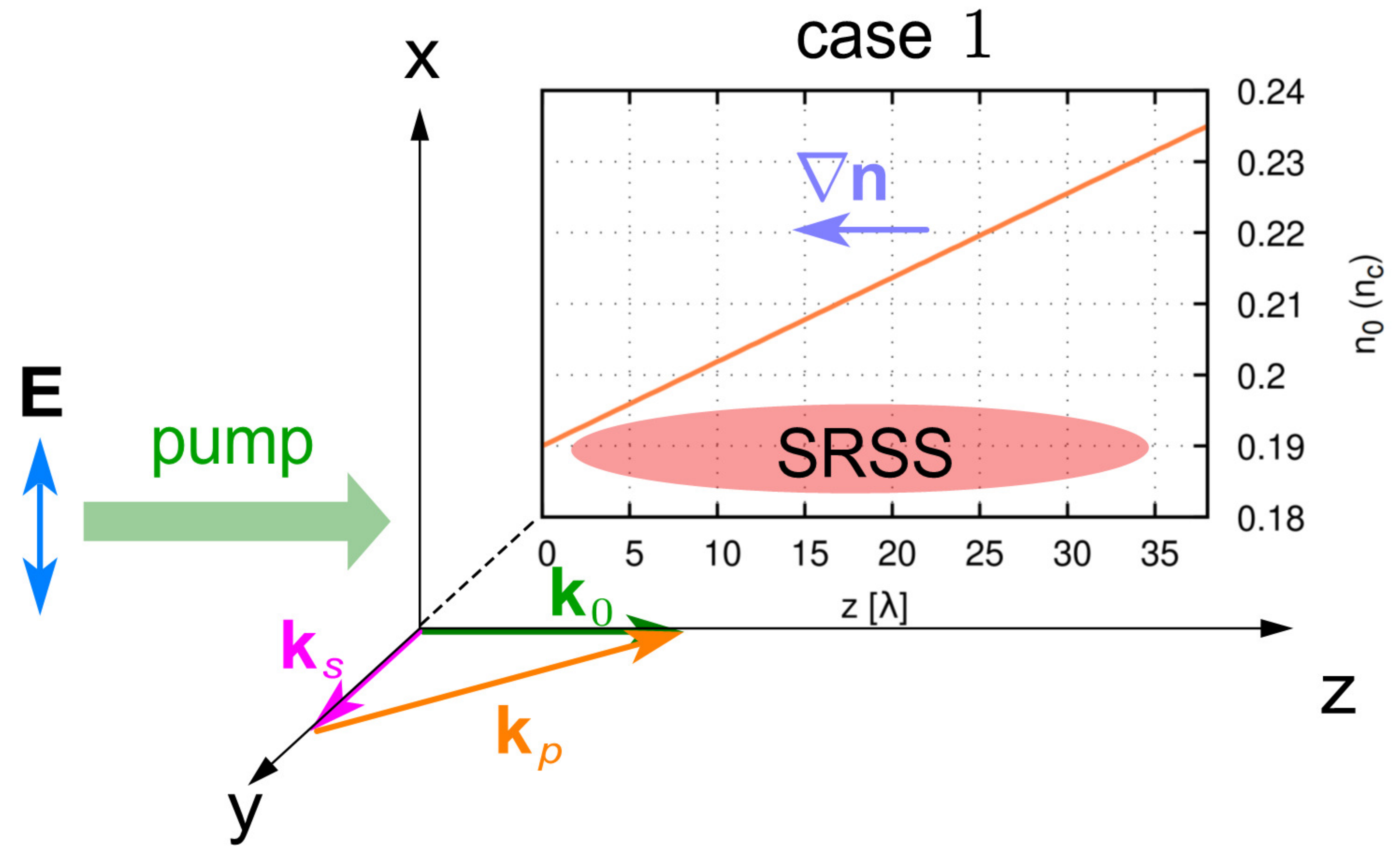}
\caption{\label{fig2} Illustration of the wave vector matching relation of an x-polarized pump in SRSS and the plasma density profile. The scattered light is marked by subscript 's' and the mageta color, the pump light is marked by the subscript '0' and the green color, while the plasma wave is marked by subscript 'p' and the orange color.  }
\end{figure}

In the following contrastive cases, except for the polarization distribution, all the other initial conditions of linearly polarized Gaussian light (case 1) and vector light (case 2) are the same. 
The long pump pulse featured a Gaussian intensity profile with a wavelength of $\lambda_0=1~\si{\mu m}$, a waist $w_0$ of $10.875\lambda_0$, and the maximum intensity of $I_0={4\times 10^{15}}~\si{W/cm^2}$, propagating along the z-direction and polarizing in the x-direction. In PIC code, the vector light is constructed by a coherent superposition of two beams with orthogonal polarization\cite{VLconstruction} and different phase distributions. In case 2, the Gaussian light polarized in the x-direction with beam waist $w_1=5\lambda_0$ and the Laguerre-Gaussian light\cite{Padgett2017OrbitalAM}, which is polarized in the y-direction and has a waist of $w_2=6\lambda_0$ and two indexes $l=1, p=0$, are coherently superposed to achieve an equivalent beam waist $w_0=10.875\lambda_0$. The superposed electric field $\mathbf{E}$ in cylindrical coordinates reads
\begin{eqnarray}\label{vector light}
\mathbf{E}&=&2\rm{exp}\left(-\frac{r^2}{w_1^2}\right)\hat{\mathbf{e}}_x\nonumber\\
~&+&1.5\left(\frac{2e}{|l|}\right)^{|l|/2}\left(\frac{r}{w_2}\right)^{|l|}\rm{exp}\left(-\frac{r^2}{w_2^2}\right)\mathrm{exp}^{il\phi}\hat{\mathbf{e}}_y,
\end{eqnarray}
where $e\approx 2.718$ is the base of the natural logarithm.
The intensity and polarization distributions of the vector light are shown in Fig.\ref{fig1}. To account for the degree of non-uniform polarization of vector light, the characteristic length of polarization variation, $L_p$, which is the distance from the position of purely x-polarized to that of purely y-polarized (or vice versus) in the cross-section of the vector light, is adopted. Owing to the high cost of three-dimensional PIC simulation, the $L_p$ of our simulation setups shown in Fig.\ref{fig2} is about ten wavelengths.
In practice, the magnitude of $L_p$ can be designed by the method outlined in Sec I.

\subsection{Simulation setup}

To isolate the effect of non-uniform polarization of vector light on SRSS and exclude the competition\cite{wenThreedimensionalParticleincellModeling2019} or transition\cite{xiaoTransitionTwoplasmonDecay2020} between different LPI processes, the physical parameters were chosen as follows. The transversely uniform hydrogen plasma density varies from $0.19n_c (z=0~\si{\mu m})$ to $0.235n_c (z=38~\si{\mu m})$ along the z-direction, corresponding to a density scale length of $L_n\approx 204~\si{\mu m}$. The electron temperature is $500~\si{eV}$, and the ion temperature is $100~\si{eV}$. The ions are set immobile. The density is chosen below the quarter-critical density because the absolute mode of TPD and SRS grow near the quarter-critical density, and they can compete\cite{wenThreedimensionalParticleincellModeling2019} or transition\cite{xiaoTransitionTwoplasmonDecay2020}, or even induce further secondary instabilities\cite{panTwoplasmonDecayInstability2018}. The convective mode of TPD grows at a relatively lower density \cite{RuiYanTPDPOP} (approximately $0.21n_c$ to $0.245n_c$). In this simulation, the short scale length makes SRBS less important, the density range excludes the absolute mode of TPD and reduces the convectively growing length of TPD, and the immobile ion eliminates the SBS. Hence SRSS is the dominant LPI process.

The simulation region measures $34~\si{\mu m}\times34~\si{\mu m}\times38~\si{\mu m}$, with a grid number of $340\times340\times 380$. Ten particles are placed in each grid, and the total simulation time is 1000 T, where T represents one laser period. Typically, periodic boundary conditions are used to facilitate the growth of SRSS\cite{wenThreedimensionalParticleincellModeling2019, xiaoStimulatedRamanSidescattering2018}. However, due to the non-periodic polarization distribution of the vector light and to ensure consistency between the two cases, open boundaries are adopted. Nonetheless, as we will demonstrate later, SRSS is rapidly excited in case 1 even with transverse open boundaries. The thermal boundary conditions are applied to particles.

\subsection{Comparison between case 1 and 2}
\begin{figure}[htpb]   
    \includegraphics[width=0.99\linewidth]{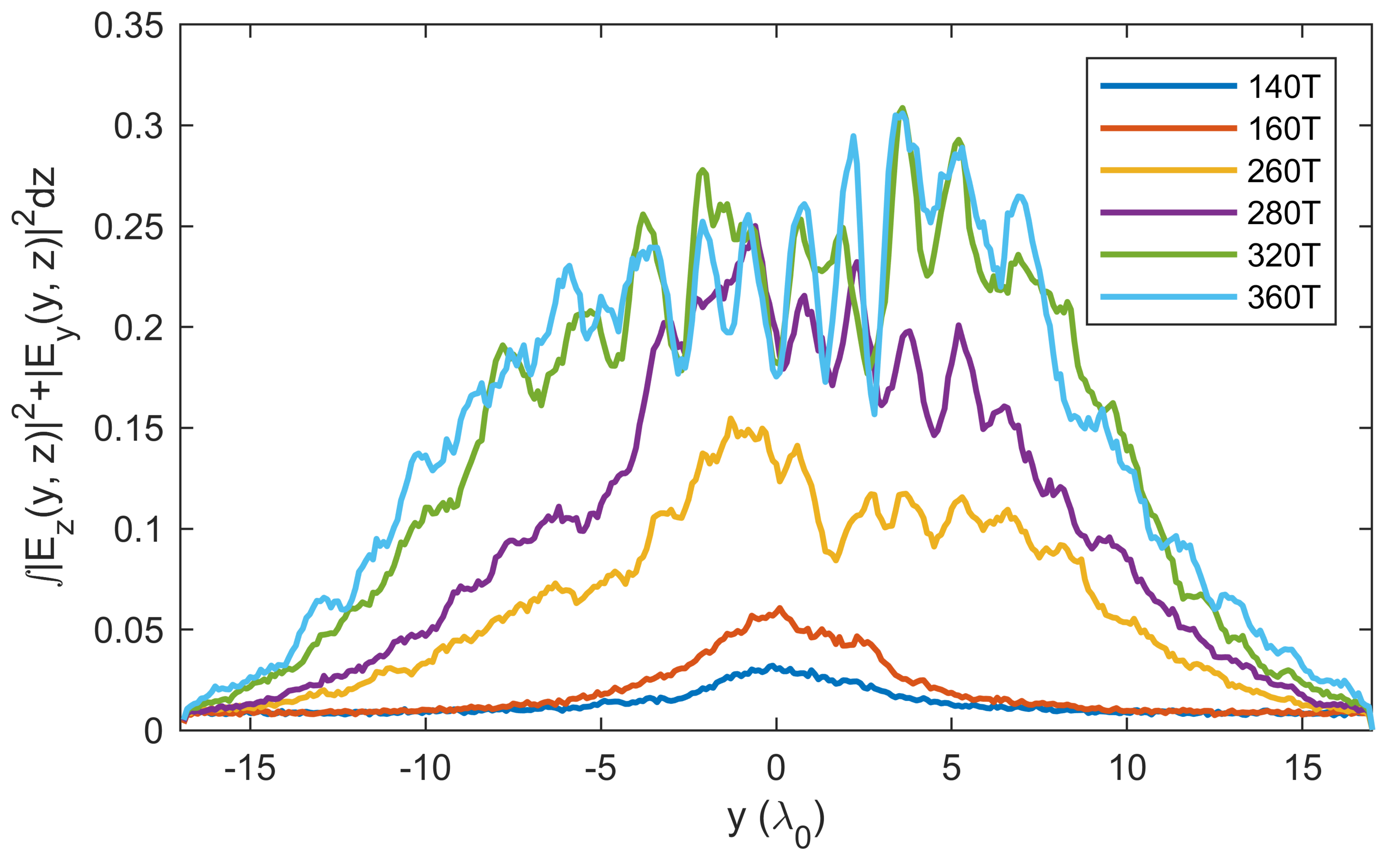}
    \caption{\label{fig3} The total energy of EPW integrated over z-direction at 140 T, 160 T, 240 T, 280 T, 320 T, and 360 T in case 1.}
    \end{figure}
\begin{figure}[htpb]  
\centering 
\includegraphics[width=0.98\linewidth]{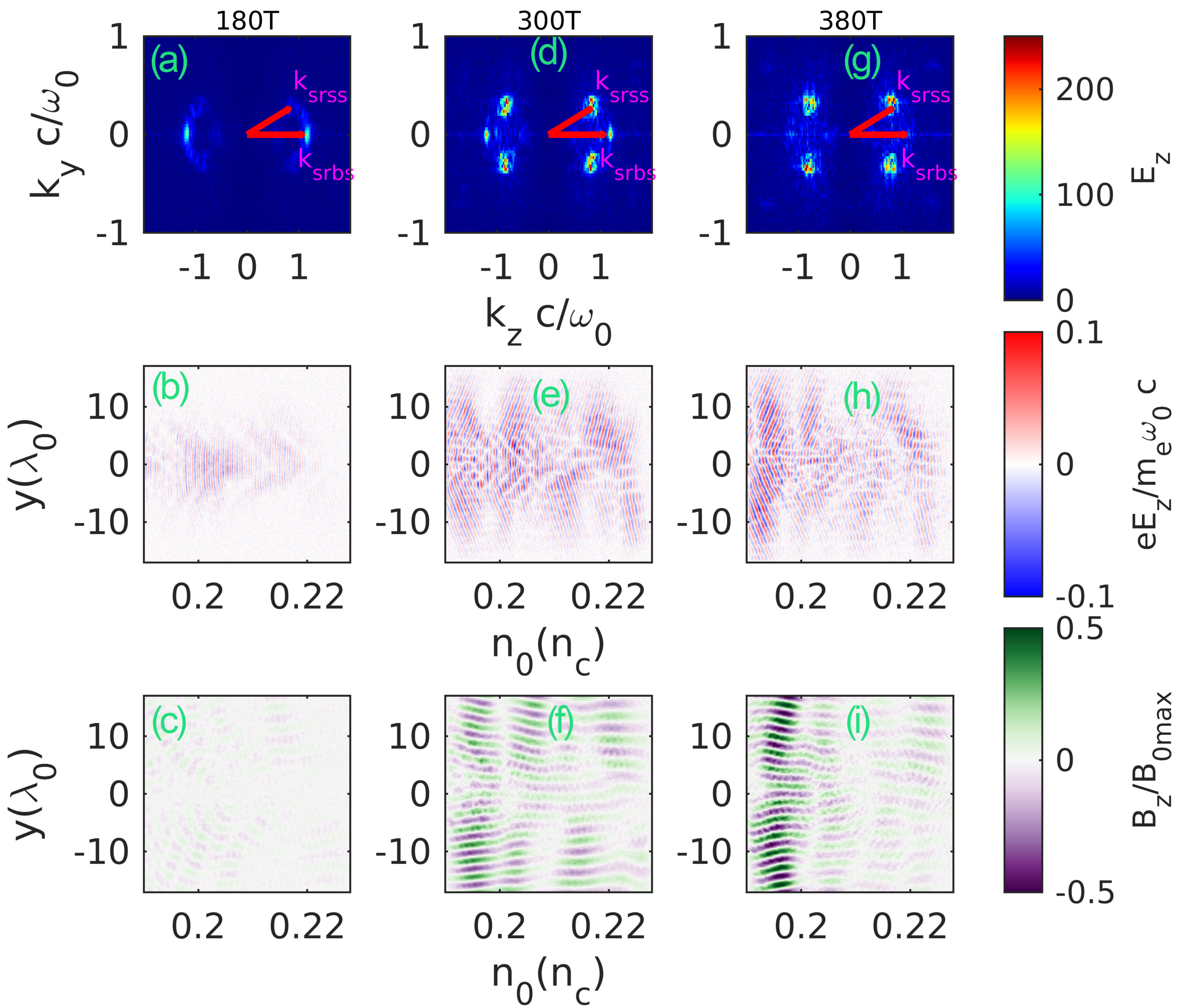}
\caption{\label{fig4} The results of case 1 in the y-z plane. The frequency spectrum in k-space of the electrostatic field $E_z(k_y,k_z)$, electrostatic field $E_z(y,z)$, and the magnetic field of scattered light of SRSS $B_z(y,z)$ at 180 T (a-c), 300 T (d-f) and 380 T (g-i).}
\end{figure}

The results of case 1 (linearly polarized Gaussian light) are shown in Fig.\ref{fig3} and \ref{fig4}. As is shown in Fig.\ref{fig2}, in case 1, the pump laser is polarized in the x-direction, the scattered wave vector at 90-degree relative to the density profile is in the y-direction, and thus the magnetic field of scattered light is $B_z$. The electrostatic wave has wave vectors (or electric fields) $k_y (E_y)$ and $k_z (E_z)$. Figure \ref{fig3} depicts the electrostatic energy integrated over z in the y-z plane, i.e., $\int{|E_z(y,z)^2+|E_y(y,z)|^2}dz$, which reflects the convective growth of SRSS in the transverse direction. The SRSS seed originates at the center owing to the maximum laser intensity, and as its amplitude grows, the profile expands and appears the flat-topped profile at about 320 T due to convective saturation.

\begin{figure*}[htpb]   
    \includegraphics[width=\linewidth]{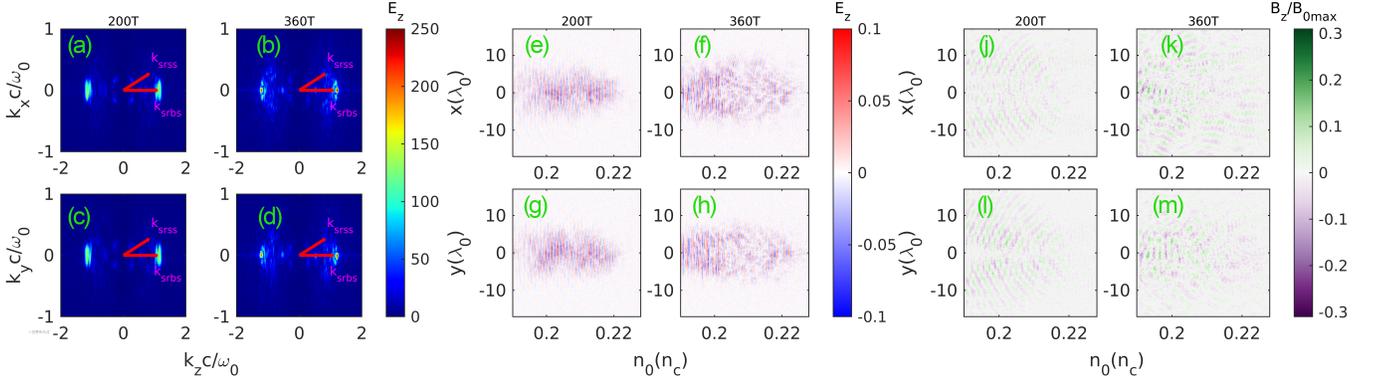}
    \caption{\label{fig5} The results of case 2. Subgraphs (a-d) are the frequency spectrum in k-space of the electrostatic field $E_z(k_y,k_z)$ and $E_z(k_x,k_z)$, (e-h) are the electrostatic field distribution $E_z(y,z)$ and $E_z(x,z)$, and (j-m) are the magnetic field of scattered light of SRSS $B_z(y,z)$ and $B_z(x,z)$ at 200 T and 360 T.}
    \end{figure*}

The electrostatic field and its k-space distribution in the occurrence plane (y-z plane) of SRSS and the magnetic field of scattered light in case 1 at different times are presented in Fig.\ref{fig4}. The three time points we have chosen are based on the convective growth of SRSS in Fig.\ref{fig3}.
The red arrows in Fig.\ref{fig4}(a), (d), and (g) denote the wave vector of the electrostatic field excited by SRSS with $\textbf{k}_{srss}=(\sqrt{1-n/n_c},\sqrt{1-2\sqrt{n/n_c}})\omega_c/c$ and SRBS with $\textbf{k}_{srbs}=(\sqrt{1-n/n_c}+\sqrt{1-2\sqrt{n/n_c}},0)\omega_0/c$, where the plasma density takes the average density $n=0.2125n_c$.
At t=180 T, the electrostatic waves were locally excited in various directions, among which the backscatter had relatively greater strength due to its higher growth rate. As the instability no longer locally develops, the intensity of SRSS is much greater than that of SRBS. At 300 T, SRSS dominates, and at 380 T, SRBS completely disappears, leaving only SRSS. 
Although periodic boundary conditions were not applied, SRSS had already convected transversely beyond the pump waist width.

Figure \ref{fig4}(c), (f), and (i) show the distribution of the magnetic field $B_z$ of scattered electromagnetic wave, which is normalized by the maximum magnetic field $B_{0max}$ of the incident laser field. At 300 T, the scattered light intensity is nearly uniform in the z-direction, while at 380 T, the scattered light is stronger near the left boundary. Despite a higher growth rate in the high–density region, the incident light enters from the left, and scatters off some energy in the low-density region, which causes an intensity distribution of the scattered light in the z-direction. Notably, the transverse intensity of the scattered light at 380 T remains significant, approximately 0.6$B_{0max}$, even without implementing periodic boundary conditions. Moreover, it is clearly shown that the scattered light presents a 90-degree scattering angle. 

    Figure \ref{fig5} provides a comparison to Fig.\ref{fig4} by illustrating the electrostatic wave and electromagnetic wave excited by vector light in case 2. The scattered wave vector of SRSS is not only in the y-direction but in all directions in the transverse (x-y) plane. Here we choose the x-z and y-z longitudinal planes to illustrate the SRS of vector light.
    The wave vector spectrum in the $(k_x, k_z)$ and $(k_y, k_z)$ planes of the electrostatic field $E_z$ are depicted in Fig.\ref{fig5} (a-d). Figure \ref{fig5} (e-h) picture the electrostatic field distribution $E_z$ and Fig.\ref{fig5} (j-m) show the magnetic field of scattered light of SRSS $B_z$ in the two longitudinal planes.
    
    At 200 T, different from case 1, the vector light only shows a backscattering signal. At 360 T, SRBS completely disappears and SRSS survives with rather high intensity in case 1, while in case 2, there is still no SRSS but with SRBS. This is because the SRSS is preferential to grow under the designed parameters, and in case 1, the SRBS is limited by the small density length scale as well as the competition of SRSS. However, in case 2, the transverse polarization variation poses a challenge for SRSS to develop, which reduces the competition and thus causes the SRBS to grow, but the amplitude of the EPW excited by SRBS is still lower than that excited by SRSS in case 1. Furthermore, Fig.\ref{fig5} (e-h) demonstrate that the width of electrostatic wave distribution is restricted within the beam waist.

    The magnetic field of the back-scattered light lies in the x-y plane and only the side-scattered signal is presented in Fig.\ref{fig5}(j-m). From $B_z$ we can distinguish a weak SRSS at 200 T. At 360 T, the intensity of the scattered electromagnetic wave is still weak. Furthermore, at 360 T, the scattering occurs in all directions, rather than at an angle of 90 degrees in case 1. 
   This also is in line with our starting point and indicates that the change in polarization direction can cause the wave vector of the scattered light to change, thereby preventing the SRSS from transverse convection.
   
\section{\label{sec3}Analysis}
The suppressive effect of vector light on SRSS has been demonstrated in Sec \ref{sec2}. In this section, the parameters design of vector light to inhibit SRSS is given.

The theory of SRSS\cite{afeyanStimulatedRamanSidescattering1985,mostromRamanSideScatterInstability1979} has been reviewed in Ref[7]. Two approaches have been developed to describe the growth of SRSS: the wave packet solution and eigenmode theory\cite{afeyanStimulatedRamanSidescattering1985,xiaoStimulatedRamanSidescattering2018}. The minimum saturation time $t_s$ of the convective mode is used as the criterion to identify whether the convective or the absolute growth dominates. If the growth time $t$ is less than $2t_s$, the wave packet (convective growth) dominates, and it has a finite exponential gain due to the refraction of scattered light in a non-uniform plasma. The minimum saturation time is\cite{xiaoStimulatedRamanSidescattering2018}
\begin{equation}
t_s\approx 8.5\left(\frac{\omega_0-\omega_p}{\omega_p}\right)^{1/2}\left(\frac{k_pv_0}{\omega_p}\right)^{1/2}\frac{L_n}{c},\label{ts}
\end{equation}
where $\omega_p$ and $\omega_0$ are the plasma frequency and laser frequency, $v_0$ and $c$ are the electron quiver velocity in laser field and the speed of light respectively, $k_p$ is the wave vector of EPW, and $L_n$ represents the density scale length. After $t>2t_s$, the eigenmode (absolute growth) dominates, and the growth rate can be found in Ref[7].
Besides, the threshold parameter is given by\cite{xiaoStimulatedRamanSidescattering2018}
\begin{equation}\label{eta}
\eta_a=(v_0/c)^{1/2}(\omega_0L_n/c)^{4/3}\frac{2-2\omega_p/\omega_0-\omega_p^2/\omega_0^2}{(\omega_p/\omega_0)^{2/3}},
\end{equation}
and the threshold is $\eta_a=1$.

Under the condition of simulation cases 1 and 2, the threshold $\eta_a$ is 58, which far exceeds the threshold. Therefore, the threshold condition is only appropriate for the pump with uniform polarization. Once there is polarization distribution in the cross-section of the pump beam, an additional threshold defined by the characteristic length of polarization variation $L_p$ should be derived. Assuming that the polarization of the initial scattered seed excited at a certain location is aligned with the polarization of the pump locally, as the scattered light convects out of the resonant location, the electric field of the pump is non-collinear with the seed. Consequently, the seed can not be amplified further and it will be damped due to Landau damping and (or) collisional damping. The condition for which the seed could grow continuously is that the growing distance $L$ before saturation is much less than $L_p$, namely,
\begin{equation}\label{threshold0}
    L=2V_{ge}t_s<L_p,
    \end{equation}
where $V_{ge}=(V_s+V_e)/2\approx V_s/2$ is the effective group velocity\cite{mostromRamanSideScatterInstability1979}, $V_e=\frac{3v_e^2k_{\perp}}{\omega_p}$ and  $V_s=\frac{c^2k_s}{\omega_0-\omega_p}$ represent the group velocity of the electrostatic wave and scattered electromagnetic wave respectively, $k_{\perp}$ is the perpendicular wave vector of the electrostatic wave, $v_e$ is the thermal velocity of electrons, and $k_s$ denotes the wave vector of scattered light. 

Substituting the saturation time gives the threshold condition defined by $L_p$
\begin{equation}\label{threshold}
8.5\left(\frac{k_pv_0}{\omega_0-\omega_p}\right)^{1/2}L_n\frac{ck_s}{\omega_p}<L_p.
\end{equation}
For the normally incident pump light, the threshold condition is
\begin{equation}\label{threshold2}
L_p^2>0.616\sqrt{I_{14}}\lambda_{\mu m}L_n^2\frac{(1-2\omega_p/\omega_0)\sqrt{2-2\omega_p/\omega_0-\omega_p^2/\omega_0^2}}{(1-\omega_p/\omega_0)\omega_p^2/\omega_0^2},
\end{equation}
where $I_{14}$ denotes the laser intensity in units of $10^{14}~\si{W/cm^2}$ and $\lambda_{\mu m}$ is the wavelength of pump light in units of the micrometer. Taking a typical parameter of ICF experiment on NIF\cite{PhysRevLett.120.055001} as an example, i.e., $I_{14}=8, \lambda_{\mu m}=0.351, L_n=500~\si{\mu m}, \omega_p/\omega_0=0.44$, where the density is about 0.2 times the critical density, the minimum characteristic length of polarization is about $0.5L_n$, which means that the SRSS can be inhibited if $L_p<0.5L_n$ even when the $\eta_a=4.6$ exceeds the threshold.

\section{\label{sec4}Discussions}

As shown by the simulations and analysis, the transverse non-uniform polarization distribution of VL contributes to the suppression of the convective growth of SRSS, which has transverse propagation and growth. It is straightforward to conjecture that the transverse non-uniform polarization has little effect on the longitudinal propagating SBRS. As is also evidenced in Fig. 5, the polarization distribution can suppress transverse convection but cannot inhibit the backscattering. 
As for TPD, considering the wave vectors of the two daughter EPWs own the transverse components, there's also an effective damping rate due to polarization change as $\nu=V_{ge}/L_p$, where $V_{ge}=V_e=\frac{3v_e^2k_{\perp}}{\omega_p}$. 
For a rough estimation, considering the plasma parameters are $T_e=2000~\si{eV}$, $L_p=15\lambda_0, n_0=0.25n_c$, and $k_{\perp}=0.1\omega_0/c, I_0=5.0\times 10^{14}~\si{W/cm^2}, \lambda=0.351~\mu m$ (the typical parameters in the ICF experiments in OMEGA), the effective damping rate is $\nu\sim 10^{-5}\omega_0$. Under these parameters, the growth rate of TPD is about $10^{-3}\omega_0$. Thus, the effective damping caused by the non-uniform polarization seems insignificant compared to the growth rate. Thus, the suppressive effect of the transverse polarization distribution on TPD is negligible because the group velocity of EPW is too small that the polarization change can not be experienced during the growth time of instability.


In this study, we propose a method to suppress SRSS using vector light, which can inhibit the convective growth of SRSS and prevent it from entering the absolute mode stage. We confirmed the effectiveness of this approach through 3D PIC simulations. Our simulation results showed that linearly polarized Gaussian light generates a strong SRSS signal in the 90-degree direction, while vector light generates few SRSS signals, even when the condition far exceeds the threshold. We also established a threshold condition based on the characteristic length of polarization variation. This work is the first to explore inhibiting laser-plasma instability from the perspective of the vector nature of the light field, offering a new possibility for suppressing LPI and paving the way for future research.

\begin{acknowledgments}
This research was supported by the Strategic Priority Research Program of Chinese Academy of Sciences, Grant No. XDA25050400 and XDA25010200, by National Natural Science Foundation of China (NSFC) under Grant No. 12175229 and 11975014.
\end{acknowledgments}

\bibliography{ref}

\begin{thebibliography}{35}%
\makeatletter
\providecommand \@ifxundefined [1]{%
 \@ifx{#1\undefined}
}%
\providecommand \@ifnum [1]{%
 \ifnum #1\expandafter \@firstoftwo
 \else \expandafter \@secondoftwo
 \fi
}%
\providecommand \@ifx [1]{%
 \ifx #1\expandafter \@firstoftwo
 \else \expandafter \@secondoftwo
 \fi
}%
\providecommand \natexlab [1]{#1}%
\providecommand \enquote  [1]{``#1''}%
\providecommand \bibnamefont  [1]{#1}%
\providecommand \bibfnamefont [1]{#1}%
\providecommand \citenamefont [1]{#1}%
\providecommand \href@noop [0]{\@secondoftwo}%
\providecommand \href [0]{\begingroup \@sanitize@url \@href}%
\providecommand \@href[1]{\@@startlink{#1}\@@href}%
\providecommand \@@href[1]{\endgroup#1\@@endlink}%
\providecommand \@sanitize@url [0]{\catcode `\\12\catcode `\$12\catcode
  `\&12\catcode `\#12\catcode `\^12\catcode `\_12\catcode `\%12\relax}%
\providecommand \@@startlink[1]{}%
\providecommand \@@endlink[0]{}%
\providecommand \url  [0]{\begingroup\@sanitize@url \@url }%
\providecommand \@url [1]{\endgroup\@href {#1}{\urlprefix }}%
\providecommand \urlprefix  [0]{URL }%
\providecommand \Eprint [0]{\href }%
\providecommand \doibase [0]{http://dx.doi.org/}%
\providecommand \selectlanguage [0]{\@gobble}%
\providecommand \bibinfo  [0]{\@secondoftwo}%
\providecommand \bibfield  [0]{\@secondoftwo}%
\providecommand \translation [1]{[#1]}%
\providecommand \BibitemOpen [0]{}%
\providecommand \bibitemStop [0]{}%
\providecommand \bibitemNoStop [0]{.\EOS\space}%
\providecommand \EOS [0]{\spacefactor3000\relax}%
\providecommand \BibitemShut  [1]{\csname bibitem#1\endcsname}%
\let\auto@bib@innerbib\@empty
\bibitem [{\citenamefont {Forslund}, \citenamefont {Kindel},\ and\
  \citenamefont {Lindman}(1975)}]{forslundTheoryStimulatedScattering1975}%
  \BibitemOpen
  \bibfield  {author} {\bibinfo {author} {\bibfnamefont {D.~W.}\ \bibnamefont
  {Forslund}}, \bibinfo {author} {\bibfnamefont {J.~M.}\ \bibnamefont
  {Kindel}}, \ and\ \bibinfo {author} {\bibfnamefont {E.~L.}\ \bibnamefont
  {Lindman}},\ }\bibfield  {title} {\enquote {\bibinfo {title} {Theory of
  stimulated scattering processes in laser-irradiated plasmas},}\ }\href
  {\doibase 10.1063/1.861248} {\bibfield  {journal} {\bibinfo  {journal}
  {Physics of Fluids}\ }\textbf {\bibinfo {volume} {18}},\ \bibinfo {pages}
  {1002} (\bibinfo {year} {1975})}\BibitemShut {NoStop}%
\bibitem [{\citenamefont {Liu}(1974)}]{liuRamanBrillouinScattering1974}%
  \BibitemOpen
  \bibfield  {author} {\bibinfo {author} {\bibfnamefont {C.~S.}\ \bibnamefont
  {Liu}},\ }\bibfield  {title} {\enquote {\bibinfo {title} {Raman and
  {{Brillouin}} scattering of electromagnetic waves in inhomogeneous
  plasmas},}\ }\href {\doibase 10.1063/1.1694867} {\bibfield  {journal}
  {\bibinfo  {journal} {Physics of Fluids}\ }\textbf {\bibinfo {volume} {17}},\
  \bibinfo {pages} {1211} (\bibinfo {year} {1974})}\BibitemShut {NoStop}%
\bibitem [{\citenamefont {Myatt}\ \emph {et~al.}(2014)\citenamefont {Myatt},
  \citenamefont {Zhang}, \citenamefont {Short}, \citenamefont {Maximov},
  \citenamefont {Seka}, \citenamefont {Froula}, \citenamefont {Edgell},
  \citenamefont {Michel}, \citenamefont {Igumenshchev}, \citenamefont {Hinkel},
  \citenamefont {Michel},\ and\ \citenamefont {Moody}}]{RN295}%
  \BibitemOpen
  \bibfield  {author} {\bibinfo {author} {\bibfnamefont {J.~F.}\ \bibnamefont
  {Myatt}}, \bibinfo {author} {\bibfnamefont {J.}~\bibnamefont {Zhang}},
  \bibinfo {author} {\bibfnamefont {R.~W.}\ \bibnamefont {Short}}, \bibinfo
  {author} {\bibfnamefont {A.~V.}\ \bibnamefont {Maximov}}, \bibinfo {author}
  {\bibfnamefont {W.}~\bibnamefont {Seka}}, \bibinfo {author} {\bibfnamefont
  {D.~H.}\ \bibnamefont {Froula}}, \bibinfo {author} {\bibfnamefont {D.~H.}\
  \bibnamefont {Edgell}}, \bibinfo {author} {\bibfnamefont {D.~T.}\
  \bibnamefont {Michel}}, \bibinfo {author} {\bibfnamefont {I.~V.}\
  \bibnamefont {Igumenshchev}}, \bibinfo {author} {\bibfnamefont {D.~E.}\
  \bibnamefont {Hinkel}}, \bibinfo {author} {\bibfnamefont {P.}~\bibnamefont
  {Michel}}, \ and\ \bibinfo {author} {\bibfnamefont {J.~D.}\ \bibnamefont
  {Moody}},\ }\bibfield  {title} {\enquote {\bibinfo {title} {Multiple-beam
  laser–plasma interactions in inertial confinement fusion},}\ }\href
  {\doibase 10.1063/1.4878623} {\bibfield  {journal} {\bibinfo  {journal}
  {Physics of Plasmas}\ }\textbf {\bibinfo {volume} {21}},\ \bibinfo {pages}
  {055501} (\bibinfo {year} {2014})}\BibitemShut {NoStop}%
\bibitem [{\citenamefont {Kruer}\ \emph {et~al.}(1996)\citenamefont {Kruer},
  \citenamefont {Wilks}, \citenamefont {Afeyan},\ and\ \citenamefont
  {Kirkwood}}]{kruerEnergyTransferCrossing1996}%
  \BibitemOpen
  \bibfield  {author} {\bibinfo {author} {\bibfnamefont {W.~L.}\ \bibnamefont
  {Kruer}}, \bibinfo {author} {\bibfnamefont {S.~C.}\ \bibnamefont {Wilks}},
  \bibinfo {author} {\bibfnamefont {B.~B.}\ \bibnamefont {Afeyan}}, \ and\
  \bibinfo {author} {\bibfnamefont {R.~K.}\ \bibnamefont {Kirkwood}},\
  }\bibfield  {title} {\enquote {\bibinfo {title} {Energy transfer between
  crossing laser beams},}\ }\href {\doibase 10.1063/1.871863} {\bibfield
  {journal} {\bibinfo  {journal} {Physics of Plasmas}\ }\textbf {\bibinfo
  {volume} {3}},\ \bibinfo {pages} {382--385} (\bibinfo {year}
  {1996})}\BibitemShut {NoStop}%
\bibitem [{\citenamefont {McKinstrie}\ \emph {et~al.}(1996)\citenamefont
  {McKinstrie}, \citenamefont {Li}, \citenamefont {Giacone},\ and\
  \citenamefont {Vu}}]{mckinstrieTwoDimensionalAnalysis1996}%
  \BibitemOpen
  \bibfield  {author} {\bibinfo {author} {\bibfnamefont {C.~J.}\ \bibnamefont
  {McKinstrie}}, \bibinfo {author} {\bibfnamefont {J.~S.}\ \bibnamefont {Li}},
  \bibinfo {author} {\bibfnamefont {R.~E.}\ \bibnamefont {Giacone}}, \ and\
  \bibinfo {author} {\bibfnamefont {H.~X.}\ \bibnamefont {Vu}},\ }\bibfield
  {title} {\enquote {\bibinfo {title} {Two-dimensional analysis of the power
  transfer between crossed laser beams},}\ }\href {\doibase 10.1063/1.871721}
  {\bibfield  {journal} {\bibinfo  {journal} {Physics of Plasmas}\ }\textbf
  {\bibinfo {volume} {3}},\ \bibinfo {pages} {2686--2692} (\bibinfo {year}
  {1996})}\BibitemShut {NoStop}%
\bibitem [{\citenamefont {Mostrom}\ and\ \citenamefont
  {Kaufman}(1979)}]{mostromRamanSideScatterInstability1979}%
  \BibitemOpen
  \bibfield  {author} {\bibinfo {author} {\bibfnamefont {M.~A.}\ \bibnamefont
  {Mostrom}}\ and\ \bibinfo {author} {\bibfnamefont {A.~N.}\ \bibnamefont
  {Kaufman}},\ }\bibfield  {title} {\enquote {\bibinfo {title} {Raman
  {{Side-Scatter Instability}} in {{Nonuniform Plasma}}},}\ }\href {\doibase
  10.1103/PhysRevLett.42.644} {\bibfield  {journal} {\bibinfo  {journal}
  {Physical Review Letters}\ }\textbf {\bibinfo {volume} {42}},\ \bibinfo
  {pages} {644--647} (\bibinfo {year} {1979})}\BibitemShut {NoStop}%
\bibitem [{\citenamefont {Xiao}\ \emph {et~al.}(2018)\citenamefont {Xiao},
  \citenamefont {Zhuo}, \citenamefont {Yin}, \citenamefont {Liu}, \citenamefont
  {Zheng}, \citenamefont {Zhao},\ and\ \citenamefont
  {He}}]{xiaoStimulatedRamanSidescattering2018}%
  \BibitemOpen
  \bibfield  {author} {\bibinfo {author} {\bibfnamefont {C.~Z.}\ \bibnamefont
  {Xiao}}, \bibinfo {author} {\bibfnamefont {H.~B.}\ \bibnamefont {Zhuo}},
  \bibinfo {author} {\bibfnamefont {Y.}~\bibnamefont {Yin}}, \bibinfo {author}
  {\bibfnamefont {Z.~J.}\ \bibnamefont {Liu}}, \bibinfo {author} {\bibfnamefont
  {C.~Y.}\ \bibnamefont {Zheng}}, \bibinfo {author} {\bibfnamefont
  {Y.}~\bibnamefont {Zhao}}, \ and\ \bibinfo {author} {\bibfnamefont {X.~T.}\
  \bibnamefont {He}},\ }\bibfield  {title} {\enquote {\bibinfo {title} {On the
  stimulated {{Raman}} sidescattering in inhomogeneous plasmas: Revisit of
  linear theory and three-dimensional particle-in-cell simulations},}\ }\href
  {\doibase 10.1088/1361-6587/aa9b41} {\bibfield  {journal} {\bibinfo
  {journal} {Plasma Physics and Controlled Fusion}\ }\textbf {\bibinfo {volume}
  {60}},\ \bibinfo {pages} {025020} (\bibinfo {year} {2018})}\BibitemShut
  {NoStop}%
\bibitem [{\citenamefont {Michel}\ \emph {et~al.}(2015)\citenamefont {Michel},
  \citenamefont {Divol}, \citenamefont {Dewald}, \citenamefont {Milovich},
  \citenamefont {Hohenberger}, \citenamefont {Jones}, \citenamefont {Hopkins},
  \citenamefont {Berger}, \citenamefont {Kruer},\ and\ \citenamefont
  {Moody}}]{michelMultibeamStimulatedRaman2015}%
  \BibitemOpen
  \bibfield  {author} {\bibinfo {author} {\bibfnamefont {P.}~\bibnamefont
  {Michel}}, \bibinfo {author} {\bibfnamefont {L.}~\bibnamefont {Divol}},
  \bibinfo {author} {\bibfnamefont {E.~L.}\ \bibnamefont {Dewald}}, \bibinfo
  {author} {\bibfnamefont {J.~L.}\ \bibnamefont {Milovich}}, \bibinfo {author}
  {\bibfnamefont {M.}~\bibnamefont {Hohenberger}}, \bibinfo {author}
  {\bibfnamefont {O.~S.}\ \bibnamefont {Jones}}, \bibinfo {author}
  {\bibfnamefont {L.~B.}\ \bibnamefont {Hopkins}}, \bibinfo {author}
  {\bibfnamefont {R.~L.}\ \bibnamefont {Berger}}, \bibinfo {author}
  {\bibfnamefont {W.~L.}\ \bibnamefont {Kruer}}, \ and\ \bibinfo {author}
  {\bibfnamefont {J.~D.}\ \bibnamefont {Moody}},\ }\bibfield  {title} {\enquote
  {\bibinfo {title} {Multibeam {{Stimulated Raman Scattering}} in {{Inertial
  Confinement Fusion Conditions}}},}\ }\href {\doibase
  10.1103/PhysRevLett.115.055003} {\bibfield  {journal} {\bibinfo  {journal}
  {Physical Review Letters}\ }\textbf {\bibinfo {volume} {115}},\ \bibinfo
  {pages} {055003} (\bibinfo {year} {2015})}\BibitemShut {NoStop}%
\bibitem [{\citenamefont {Dewald}\ \emph {et~al.}(2016)\citenamefont {Dewald},
  \citenamefont {Hartemann}, \citenamefont {Michel}, \citenamefont {Milovich},
  \citenamefont {Hohenberger}, \citenamefont {Pak}, \citenamefont {Landen},
  \citenamefont {Divol}, \citenamefont {Robey}, \citenamefont {Hurricane},
  \citenamefont {D{\"o}ppner}, \citenamefont {Albert}, \citenamefont
  {Bachmann}, \citenamefont {Meezan}, \citenamefont {MacKinnon}, \citenamefont
  {Callahan},\ and\ \citenamefont {Edwards}}]{Dewald2016}%
  \BibitemOpen
  \bibfield  {author} {\bibinfo {author} {\bibfnamefont {E.~L.}\ \bibnamefont
  {Dewald}}, \bibinfo {author} {\bibfnamefont {F.}~\bibnamefont {Hartemann}},
  \bibinfo {author} {\bibfnamefont {P.}~\bibnamefont {Michel}}, \bibinfo
  {author} {\bibfnamefont {J.}~\bibnamefont {Milovich}}, \bibinfo {author}
  {\bibfnamefont {M.}~\bibnamefont {Hohenberger}}, \bibinfo {author}
  {\bibfnamefont {A.}~\bibnamefont {Pak}}, \bibinfo {author} {\bibfnamefont
  {O.~L.}\ \bibnamefont {Landen}}, \bibinfo {author} {\bibfnamefont
  {L.}~\bibnamefont {Divol}}, \bibinfo {author} {\bibfnamefont {H.~F.}\
  \bibnamefont {Robey}}, \bibinfo {author} {\bibfnamefont {O.~A.}\ \bibnamefont
  {Hurricane}}, \bibinfo {author} {\bibfnamefont {T.}~\bibnamefont
  {D{\"o}ppner}}, \bibinfo {author} {\bibfnamefont {F.}~\bibnamefont {Albert}},
  \bibinfo {author} {\bibfnamefont {B.}~\bibnamefont {Bachmann}}, \bibinfo
  {author} {\bibfnamefont {N.~B.}\ \bibnamefont {Meezan}}, \bibinfo {author}
  {\bibfnamefont {A.~J.}\ \bibnamefont {MacKinnon}}, \bibinfo {author}
  {\bibfnamefont {D.}~\bibnamefont {Callahan}}, \ and\ \bibinfo {author}
  {\bibfnamefont {M.~J.}\ \bibnamefont {Edwards}},\ }\bibfield  {title}
  {\enquote {\bibinfo {title} {Generation and beaming of early hot electrons
  onto the capsule in laser-driven ignition hohlraums},}\ }\href {\doibase
  10.1103/PhysRevLett.116.075003} {\bibfield  {journal} {\bibinfo  {journal}
  {Physical Review Letters}\ }\textbf {\bibinfo {volume} {116}},\ \bibinfo
  {pages} {075003} (\bibinfo {year} {2016})}\BibitemShut {NoStop}%
\bibitem [{\citenamefont {Craxton}\ \emph {et~al.}(2015)\citenamefont
  {Craxton}, \citenamefont {Anderson}, \citenamefont {Boehly}, \citenamefont
  {Goncharov}, \citenamefont {Harding}, \citenamefont {Knauer}, \citenamefont
  {McCrory}, \citenamefont {McKenty}, \citenamefont {Meyerhofer}, \citenamefont
  {Myatt}, \citenamefont {Schmitt}, \citenamefont {Sethian}, \citenamefont
  {Short}, \citenamefont {Skupsky}, \citenamefont {Theobald}, \citenamefont
  {Kruer}, \citenamefont {Tanaka}, \citenamefont {Betti}, \citenamefont
  {Collins}, \citenamefont {Delettrez}, \citenamefont {Hu}, \citenamefont
  {Marozas}, \citenamefont {Maximov}, \citenamefont {Michel}, \citenamefont
  {Radha}, \citenamefont {Regan}, \citenamefont {Sangster}, \citenamefont
  {Seka}, \citenamefont {Solodov}, \citenamefont {Soures}, \citenamefont
  {Stoeckl},\ and\ \citenamefont {Zuegel}}]{DDreview2015}%
  \BibitemOpen
  \bibfield  {author} {\bibinfo {author} {\bibfnamefont {R.~S.}\ \bibnamefont
  {Craxton}}, \bibinfo {author} {\bibfnamefont {K.~S.}\ \bibnamefont
  {Anderson}}, \bibinfo {author} {\bibfnamefont {T.~R.}\ \bibnamefont
  {Boehly}}, \bibinfo {author} {\bibfnamefont {V.~N.}\ \bibnamefont
  {Goncharov}}, \bibinfo {author} {\bibfnamefont {D.~R.}\ \bibnamefont
  {Harding}}, \bibinfo {author} {\bibfnamefont {J.~P.}\ \bibnamefont {Knauer}},
  \bibinfo {author} {\bibfnamefont {R.~L.}\ \bibnamefont {McCrory}}, \bibinfo
  {author} {\bibfnamefont {P.~W.}\ \bibnamefont {McKenty}}, \bibinfo {author}
  {\bibfnamefont {D.~D.}\ \bibnamefont {Meyerhofer}}, \bibinfo {author}
  {\bibfnamefont {J.~F.}\ \bibnamefont {Myatt}}, \bibinfo {author}
  {\bibfnamefont {A.~J.}\ \bibnamefont {Schmitt}}, \bibinfo {author}
  {\bibfnamefont {J.~D.}\ \bibnamefont {Sethian}}, \bibinfo {author}
  {\bibfnamefont {R.~W.}\ \bibnamefont {Short}}, \bibinfo {author}
  {\bibfnamefont {S.}~\bibnamefont {Skupsky}}, \bibinfo {author} {\bibfnamefont
  {W.}~\bibnamefont {Theobald}}, \bibinfo {author} {\bibfnamefont {W.~L.}\
  \bibnamefont {Kruer}}, \bibinfo {author} {\bibfnamefont {K.}~\bibnamefont
  {Tanaka}}, \bibinfo {author} {\bibfnamefont {R.}~\bibnamefont {Betti}},
  \bibinfo {author} {\bibfnamefont {T.~J.~B.}\ \bibnamefont {Collins}},
  \bibinfo {author} {\bibfnamefont {J.~A.}\ \bibnamefont {Delettrez}}, \bibinfo
  {author} {\bibfnamefont {S.~X.}\ \bibnamefont {Hu}}, \bibinfo {author}
  {\bibfnamefont {J.~A.}\ \bibnamefont {Marozas}}, \bibinfo {author}
  {\bibfnamefont {A.~V.}\ \bibnamefont {Maximov}}, \bibinfo {author}
  {\bibfnamefont {D.~T.}\ \bibnamefont {Michel}}, \bibinfo {author}
  {\bibfnamefont {P.~B.}\ \bibnamefont {Radha}}, \bibinfo {author}
  {\bibfnamefont {S.~P.}\ \bibnamefont {Regan}}, \bibinfo {author}
  {\bibfnamefont {T.~C.}\ \bibnamefont {Sangster}}, \bibinfo {author}
  {\bibfnamefont {W.}~\bibnamefont {Seka}}, \bibinfo {author} {\bibfnamefont
  {A.~A.}\ \bibnamefont {Solodov}}, \bibinfo {author} {\bibfnamefont {J.~M.}\
  \bibnamefont {Soures}}, \bibinfo {author} {\bibfnamefont {C.}~\bibnamefont
  {Stoeckl}}, \ and\ \bibinfo {author} {\bibfnamefont {J.~D.}\ \bibnamefont
  {Zuegel}},\ }\bibfield  {title} {\enquote {\bibinfo {title} {{Direct-drive
  inertial confinement fusion: A review}},}\ }\href {\doibase
  10.1063/1.4934714} {\bibfield  {journal} {\bibinfo  {journal} {Physics of
  Plasmas}\ }\textbf {\bibinfo {volume} {22}},\ \bibinfo {pages} {110501}
  (\bibinfo {year} {2015})}\BibitemShut {NoStop}%
\bibitem [{\citenamefont {Betti}\ \emph {et~al.}(2007)\citenamefont {Betti},
  \citenamefont {Zhou}, \citenamefont {Anderson}, \citenamefont {Perkins},
  \citenamefont {Theobald},\ and\ \citenamefont
  {Solodov}}]{PhysRevLett.98.155001}%
  \BibitemOpen
  \bibfield  {author} {\bibinfo {author} {\bibfnamefont {R.}~\bibnamefont
  {Betti}}, \bibinfo {author} {\bibfnamefont {C.~D.}\ \bibnamefont {Zhou}},
  \bibinfo {author} {\bibfnamefont {K.~S.}\ \bibnamefont {Anderson}}, \bibinfo
  {author} {\bibfnamefont {L.~J.}\ \bibnamefont {Perkins}}, \bibinfo {author}
  {\bibfnamefont {W.}~\bibnamefont {Theobald}}, \ and\ \bibinfo {author}
  {\bibfnamefont {A.~A.}\ \bibnamefont {Solodov}},\ }\bibfield  {title}
  {\enquote {\bibinfo {title} {Shock ignition of thermonuclear fuel with high
  areal density},}\ }\href {\doibase 10.1103/PhysRevLett.98.155001} {\bibfield
  {journal} {\bibinfo  {journal} {Phys. Rev. Lett.}\ }\textbf {\bibinfo
  {volume} {98}},\ \bibinfo {pages} {155001} (\bibinfo {year}
  {2007})}\BibitemShut {NoStop}%
\bibitem [{\citenamefont {Depierreux}\ \emph {et~al.}(2016)\citenamefont
  {Depierreux}, \citenamefont {Neuville}, \citenamefont {Baccou}, \citenamefont
  {Tassin}, \citenamefont {Casanova}, \citenamefont {{Masson-Laborde}},
  \citenamefont {Borisenko}, \citenamefont {Orekhov}, \citenamefont {Colaitis},
  \citenamefont {Debayle}, \citenamefont {Duchateau}, \citenamefont {Heron},
  \citenamefont {Huller}, \citenamefont {Loiseau}, \citenamefont {Nicola{\"i}},
  \citenamefont {Pesme}, \citenamefont {Riconda}, \citenamefont {Tran},
  \citenamefont {Bahr}, \citenamefont {Katz}, \citenamefont {Stoeckl},
  \citenamefont {Seka}, \citenamefont {Tikhonchuk},\ and\ \citenamefont
  {Labaune}}]{PhysRevLett.117.235002}%
  \BibitemOpen
  \bibfield  {author} {\bibinfo {author} {\bibfnamefont {S.}~\bibnamefont
  {Depierreux}}, \bibinfo {author} {\bibfnamefont {C.}~\bibnamefont
  {Neuville}}, \bibinfo {author} {\bibfnamefont {C.}~\bibnamefont {Baccou}},
  \bibinfo {author} {\bibfnamefont {V.}~\bibnamefont {Tassin}}, \bibinfo
  {author} {\bibfnamefont {M.}~\bibnamefont {Casanova}}, \bibinfo {author}
  {\bibfnamefont {P.-E.}\ \bibnamefont {{Masson-Laborde}}}, \bibinfo {author}
  {\bibfnamefont {N.}~\bibnamefont {Borisenko}}, \bibinfo {author}
  {\bibfnamefont {A.}~\bibnamefont {Orekhov}}, \bibinfo {author} {\bibfnamefont
  {A.}~\bibnamefont {Colaitis}}, \bibinfo {author} {\bibfnamefont
  {A.}~\bibnamefont {Debayle}}, \bibinfo {author} {\bibfnamefont
  {G.}~\bibnamefont {Duchateau}}, \bibinfo {author} {\bibfnamefont
  {A.}~\bibnamefont {Heron}}, \bibinfo {author} {\bibfnamefont
  {S.}~\bibnamefont {Huller}}, \bibinfo {author} {\bibfnamefont
  {P.}~\bibnamefont {Loiseau}}, \bibinfo {author} {\bibfnamefont
  {P.}~\bibnamefont {Nicola{\"i}}}, \bibinfo {author} {\bibfnamefont
  {D.}~\bibnamefont {Pesme}}, \bibinfo {author} {\bibfnamefont
  {C.}~\bibnamefont {Riconda}}, \bibinfo {author} {\bibfnamefont
  {G.}~\bibnamefont {Tran}}, \bibinfo {author} {\bibfnamefont {R.}~\bibnamefont
  {Bahr}}, \bibinfo {author} {\bibfnamefont {J.}~\bibnamefont {Katz}}, \bibinfo
  {author} {\bibfnamefont {C.}~\bibnamefont {Stoeckl}}, \bibinfo {author}
  {\bibfnamefont {W.}~\bibnamefont {Seka}}, \bibinfo {author} {\bibfnamefont
  {V.}~\bibnamefont {Tikhonchuk}}, \ and\ \bibinfo {author} {\bibfnamefont
  {C.}~\bibnamefont {Labaune}},\ }\bibfield  {title} {\enquote {\bibinfo
  {title} {Experimental investigation of the collective raman scattering of
  multiple laser beams in inhomogeneous plasmas},}\ }\href {\doibase
  10.1103/PhysRevLett.117.235002} {\bibfield  {journal} {\bibinfo  {journal}
  {Physical Review Letters}\ }\textbf {\bibinfo {volume} {117}},\ \bibinfo
  {pages} {235002} (\bibinfo {year} {2016})}\BibitemShut {NoStop}%
\bibitem [{\citenamefont {Rosenberg}\ \emph {et~al.}(2018)\citenamefont
  {Rosenberg}, \citenamefont {Solodov}, \citenamefont {Myatt}, \citenamefont
  {Seka}, \citenamefont {Michel}, \citenamefont {Hohenberger}, \citenamefont
  {Short}, \citenamefont {Epstein}, \citenamefont {Regan}, \citenamefont
  {Campbell}, \citenamefont {Chapman}, \citenamefont {Goyon}, \citenamefont
  {Ralph}, \citenamefont {Barrios}, \citenamefont {Moody},\ and\ \citenamefont
  {Bates}}]{PhysRevLett.120.055001}%
  \BibitemOpen
  \bibfield  {author} {\bibinfo {author} {\bibfnamefont {M.~J.}\ \bibnamefont
  {Rosenberg}}, \bibinfo {author} {\bibfnamefont {A.~A.}\ \bibnamefont
  {Solodov}}, \bibinfo {author} {\bibfnamefont {J.~F.}\ \bibnamefont {Myatt}},
  \bibinfo {author} {\bibfnamefont {W.}~\bibnamefont {Seka}}, \bibinfo {author}
  {\bibfnamefont {P.}~\bibnamefont {Michel}}, \bibinfo {author} {\bibfnamefont
  {M.}~\bibnamefont {Hohenberger}}, \bibinfo {author} {\bibfnamefont {R.~W.}\
  \bibnamefont {Short}}, \bibinfo {author} {\bibfnamefont {R.}~\bibnamefont
  {Epstein}}, \bibinfo {author} {\bibfnamefont {S.~P.}\ \bibnamefont {Regan}},
  \bibinfo {author} {\bibfnamefont {E.~M.}\ \bibnamefont {Campbell}}, \bibinfo
  {author} {\bibfnamefont {T.}~\bibnamefont {Chapman}}, \bibinfo {author}
  {\bibfnamefont {C.}~\bibnamefont {Goyon}}, \bibinfo {author} {\bibfnamefont
  {J.~E.}\ \bibnamefont {Ralph}}, \bibinfo {author} {\bibfnamefont {M.~A.}\
  \bibnamefont {Barrios}}, \bibinfo {author} {\bibfnamefont {J.~D.}\
  \bibnamefont {Moody}}, \ and\ \bibinfo {author} {\bibfnamefont {J.~W.}\
  \bibnamefont {Bates}},\ }\bibfield  {title} {\enquote {\bibinfo {title}
  {Origins and scaling of hot-electron preheat in ignition-scale direct-drive
  inertial confinement fusion experiments},}\ }\href {\doibase
  10.1103/PhysRevLett.120.055001} {\bibfield  {journal} {\bibinfo  {journal}
  {Physical Review Letters}\ }\textbf {\bibinfo {volume} {120}},\ \bibinfo
  {pages} {055001} (\bibinfo {year} {2018})}\BibitemShut {NoStop}%
\bibitem [{\citenamefont {Michel}\ \emph {et~al.}(2019)\citenamefont {Michel},
  \citenamefont {Rosenberg}, \citenamefont {Seka}, \citenamefont {Solodov},
  \citenamefont {Short}, \citenamefont {Chapman}, \citenamefont {Goyon},
  \citenamefont {Lemos}, \citenamefont {Hohenberger}, \citenamefont {Moody},
  \citenamefont {Regan},\ and\ \citenamefont
  {Myatt}}]{michelTheoryMeasurementsConvective2019a}%
  \BibitemOpen
  \bibfield  {author} {\bibinfo {author} {\bibfnamefont {P.}~\bibnamefont
  {Michel}}, \bibinfo {author} {\bibfnamefont {M.~J.}\ \bibnamefont
  {Rosenberg}}, \bibinfo {author} {\bibfnamefont {W.}~\bibnamefont {Seka}},
  \bibinfo {author} {\bibfnamefont {A.~A.}\ \bibnamefont {Solodov}}, \bibinfo
  {author} {\bibfnamefont {R.~W.}\ \bibnamefont {Short}}, \bibinfo {author}
  {\bibfnamefont {T.}~\bibnamefont {Chapman}}, \bibinfo {author} {\bibfnamefont
  {C.}~\bibnamefont {Goyon}}, \bibinfo {author} {\bibfnamefont
  {N.}~\bibnamefont {Lemos}}, \bibinfo {author} {\bibfnamefont
  {M.}~\bibnamefont {Hohenberger}}, \bibinfo {author} {\bibfnamefont {J.~D.}\
  \bibnamefont {Moody}}, \bibinfo {author} {\bibfnamefont {S.~P.}\ \bibnamefont
  {Regan}}, \ and\ \bibinfo {author} {\bibfnamefont {J.~F.}\ \bibnamefont
  {Myatt}},\ }\bibfield  {title} {\enquote {\bibinfo {title} {Theory and
  measurements of convective {{Raman}} side scatter in inertial confinement
  fusion experiments},}\ }\href {\doibase 10.1103/PhysRevE.99.033203}
  {\bibfield  {journal} {\bibinfo  {journal} {Physical Review E}\ }\textbf
  {\bibinfo {volume} {99}},\ \bibinfo {pages} {033203} (\bibinfo {year}
  {2019})}\BibitemShut {NoStop}%
\bibitem [{\citenamefont {Cristoforetti}\ \emph {et~al.}(2019)\citenamefont
  {Cristoforetti}, \citenamefont {Antonelli}, \citenamefont {Mancelli},
  \citenamefont {Atzeni}, \citenamefont {Baffigi}, \citenamefont {Barbato},
  \citenamefont {Batani}, \citenamefont {Boutoux}, \citenamefont {D’Amato},
  \citenamefont {Dostal},\ and\ \citenamefont {et~al.}}]{cristoforetti_2019}%
  \BibitemOpen
  \bibfield  {author} {\bibinfo {author} {\bibfnamefont {G.}~\bibnamefont
  {Cristoforetti}}, \bibinfo {author} {\bibfnamefont {L.}~\bibnamefont
  {Antonelli}}, \bibinfo {author} {\bibfnamefont {D.}~\bibnamefont {Mancelli}},
  \bibinfo {author} {\bibfnamefont {S.}~\bibnamefont {Atzeni}}, \bibinfo
  {author} {\bibfnamefont {F.}~\bibnamefont {Baffigi}}, \bibinfo {author}
  {\bibfnamefont {F.}~\bibnamefont {Barbato}}, \bibinfo {author} {\bibfnamefont
  {D.}~\bibnamefont {Batani}}, \bibinfo {author} {\bibfnamefont
  {G.}~\bibnamefont {Boutoux}}, \bibinfo {author} {\bibfnamefont
  {F.}~\bibnamefont {D’Amato}}, \bibinfo {author} {\bibfnamefont
  {J.}~\bibnamefont {Dostal}}, \ and\ \bibinfo {author} {\bibnamefont
  {et~al.}},\ }\bibfield  {title} {\enquote {\bibinfo {title} {Time evolution
  of stimulated raman scattering and two-plasmon decay at laser intensities
  relevant for shock ignition in a hot plasma},}\ }\href {\doibase
  10.1017/hpl.2019.37} {\bibfield  {journal} {\bibinfo  {journal} {High Power
  Laser Science and Engineering}\ }\textbf {\bibinfo {volume} {7}},\ \bibinfo
  {pages} {e51} (\bibinfo {year} {2019})}\BibitemShut {NoStop}%
\bibitem [{\citenamefont {Glize}\ \emph {et~al.}(2023)\citenamefont {Glize},
  \citenamefont {Zhao}, \citenamefont {Zhang}, \citenamefont {Lian},
  \citenamefont {Tan}, \citenamefont {Wu}, \citenamefont {Xiao}, \citenamefont
  {Yan}, \citenamefont {Zhang}, \citenamefont {Yuan},\ and\ \citenamefont
  {Zhang}}]{glizeMeasurementStimulatedRaman2023}%
  \BibitemOpen
  \bibfield  {author} {\bibinfo {author} {\bibfnamefont {K.}~\bibnamefont
  {Glize}}, \bibinfo {author} {\bibfnamefont {X.}~\bibnamefont {Zhao}},
  \bibinfo {author} {\bibfnamefont {Y.}~\bibnamefont {Zhang}}, \bibinfo
  {author} {\bibfnamefont {C.}~\bibnamefont {Lian}}, \bibinfo {author}
  {\bibfnamefont {S.}~\bibnamefont {Tan}}, \bibinfo {author} {\bibfnamefont
  {F.}~\bibnamefont {Wu}}, \bibinfo {author} {\bibfnamefont {C.}~\bibnamefont
  {Xiao}}, \bibinfo {author} {\bibfnamefont {R.}~\bibnamefont {Yan}}, \bibinfo
  {author} {\bibfnamefont {Z.}~\bibnamefont {Zhang}}, \bibinfo {author}
  {\bibfnamefont {X.}~\bibnamefont {Yuan}}, \ and\ \bibinfo {author}
  {\bibfnamefont {J.}~\bibnamefont {Zhang}},\ }\href {\doibase
  10.48550/arXiv.2209.08251} {\enquote {\bibinfo {title} {Measurement of
  {{Stimulated Raman Side-Scattering Energetic Importance}} in {{Directly
  Driven Experiment}}},}\ } (\bibinfo {year} {2023}),\ \Eprint
  {http://arxiv.org/abs/2209.08251} {arxiv:2209.08251 [physics]} \BibitemShut
  {NoStop}%
\bibitem [{\citenamefont {Zhang}\ \emph {et~al.}(2020)\citenamefont {Zhang},
  \citenamefont {Wang}, \citenamefont {Yang}, \citenamefont {Wu}, \citenamefont
  {Ma}, \citenamefont {Jiao}, \citenamefont {Zhang}, \citenamefont {Wu},
  \citenamefont {Yuan}, \citenamefont {Li},\ and\ \citenamefont
  {Zhu}}]{zhangj2020}%
  \BibitemOpen
  \bibfield  {author} {\bibinfo {author} {\bibfnamefont {J.}~\bibnamefont
  {Zhang}}, \bibinfo {author} {\bibfnamefont {W.~M.}\ \bibnamefont {Wang}},
  \bibinfo {author} {\bibfnamefont {X.~H.}\ \bibnamefont {Yang}}, \bibinfo
  {author} {\bibfnamefont {D.}~\bibnamefont {Wu}}, \bibinfo {author}
  {\bibfnamefont {Y.~Y.}\ \bibnamefont {Ma}}, \bibinfo {author} {\bibfnamefont
  {J.~L.}\ \bibnamefont {Jiao}}, \bibinfo {author} {\bibfnamefont
  {Z.}~\bibnamefont {Zhang}}, \bibinfo {author} {\bibfnamefont {F.~Y.}\
  \bibnamefont {Wu}}, \bibinfo {author} {\bibfnamefont {X.~H.}\ \bibnamefont
  {Yuan}}, \bibinfo {author} {\bibfnamefont {Y.~T.}\ \bibnamefont {Li}}, \ and\
  \bibinfo {author} {\bibfnamefont {J.~Q.}\ \bibnamefont {Zhu}},\ }\bibfield
  {title} {\enquote {\bibinfo {title} {Double-cone ignition scheme for inertial
  confinement fusion},}\ }\href {\doibase 10.1098/rsta.2020.0015} {\bibfield
  {journal} {\bibinfo  {journal} {Philosophical Transactions of the Royal
  Society A: Mathematical, Physical and Engineering Sciences}\ }\textbf
  {\bibinfo {volume} {378}},\ \bibinfo {pages} {20200015} (\bibinfo {year}
  {2020})},\ \Eprint
  {http://arxiv.org/abs/https://royalsocietypublishing.org/doi/pdf/10.1098/rsta.2020.0015}
  {https://royalsocietypublishing.org/doi/pdf/10.1098/rsta.2020.0015}
  \BibitemShut {NoStop}%
\bibitem [{\citenamefont {Rosales-Guzm{\'a}n}, \citenamefont {Ndagano},\ and\
  \citenamefont {Forbes}(2018)}]{VLreview2018}%
  \BibitemOpen
  \bibfield  {author} {\bibinfo {author} {\bibfnamefont {C.}~\bibnamefont
  {Rosales-Guzm{\'a}n}}, \bibinfo {author} {\bibfnamefont {B.}~\bibnamefont
  {Ndagano}}, \ and\ \bibinfo {author} {\bibfnamefont {A.}~\bibnamefont
  {Forbes}},\ }\bibfield  {title} {\enquote {\bibinfo {title} {A review of
  complex vector light fields and their applications},}\ }\href@noop {}
  {\bibfield  {journal} {\bibinfo  {journal} {Journal of Optics}\ }\textbf
  {\bibinfo {volume} {20}} (\bibinfo {year} {2018})}\BibitemShut {NoStop}%
\bibitem [{\citenamefont {Zhan}(2009)}]{zhanCylindricalVectorBeams2009}%
  \BibitemOpen
  \bibfield  {author} {\bibinfo {author} {\bibfnamefont {Q.}~\bibnamefont
  {Zhan}},\ }\bibfield  {title} {\enquote {\bibinfo {title} {Cylindrical vector
  beams: From mathematical concepts to applications},}\ }\href {\doibase
  10.1364/AOP.1.000001} {\bibfield  {journal} {\bibinfo  {journal} {Advances in
  Optics and Photonics}\ }\textbf {\bibinfo {volume} {1}},\ \bibinfo {pages}
  {1} (\bibinfo {year} {2009})}\BibitemShut {NoStop}%
\bibitem [{\citenamefont {Wang}\ \emph {et~al.}(2007)\citenamefont {Wang},
  \citenamefont {Ding}, \citenamefont {Ni}, \citenamefont {Guo},\ and\
  \citenamefont {Wang}}]{wangGenerationArbitraryVector2007}%
  \BibitemOpen
  \bibfield  {author} {\bibinfo {author} {\bibfnamefont {X.-L.}\ \bibnamefont
  {Wang}}, \bibinfo {author} {\bibfnamefont {J.}~\bibnamefont {Ding}}, \bibinfo
  {author} {\bibfnamefont {W.-J.}\ \bibnamefont {Ni}}, \bibinfo {author}
  {\bibfnamefont {C.-S.}\ \bibnamefont {Guo}}, \ and\ \bibinfo {author}
  {\bibfnamefont {H.-T.}\ \bibnamefont {Wang}},\ }\bibfield  {title} {\enquote
  {\bibinfo {title} {Generation of arbitrary vector beams with a spatial light
  modulator and a common path interferometric arrangement},}\ }\href {\doibase
  10.1364/OL.32.003549} {\bibfield  {journal} {\bibinfo  {journal} {Optics
  Letters}\ }\textbf {\bibinfo {volume} {32}},\ \bibinfo {pages} {3549--3551}
  (\bibinfo {year} {2007})}\BibitemShut {NoStop}%
\bibitem [{\citenamefont {Wang}\ \emph {et~al.}(2010)\citenamefont {Wang},
  \citenamefont {Li}, \citenamefont {Chen}, \citenamefont {Guo}, \citenamefont
  {Ding},\ and\ \citenamefont {Wang}}]{wangNewTypeVector2010}%
  \BibitemOpen
  \bibfield  {author} {\bibinfo {author} {\bibfnamefont {X.-L.}\ \bibnamefont
  {Wang}}, \bibinfo {author} {\bibfnamefont {Y.}~\bibnamefont {Li}}, \bibinfo
  {author} {\bibfnamefont {J.}~\bibnamefont {Chen}}, \bibinfo {author}
  {\bibfnamefont {C.-S.}\ \bibnamefont {Guo}}, \bibinfo {author} {\bibfnamefont
  {J.}~\bibnamefont {Ding}}, \ and\ \bibinfo {author} {\bibfnamefont {H.-T.}\
  \bibnamefont {Wang}},\ }\bibfield  {title} {\enquote {\bibinfo {title} {A new
  type of vector fields with hybrid states of polarization},}\ }\href {\doibase
  10.1364/OE.18.010786} {\bibfield  {journal} {\bibinfo  {journal} {Optics
  Express}\ }\textbf {\bibinfo {volume} {18}},\ \bibinfo {pages} {10786--10795}
  (\bibinfo {year} {2010})}\BibitemShut {NoStop}%
\bibitem [{\citenamefont {Follett}\ \emph {et~al.}(2019)\citenamefont
  {Follett}, \citenamefont {Shaw}, \citenamefont {Myatt}, \citenamefont
  {Dorrer}, \citenamefont {Froula},\ and\ \citenamefont
  {Palastro}}]{follettThresholdsAbsoluteInstabilities2019}%
  \BibitemOpen
  \bibfield  {author} {\bibinfo {author} {\bibfnamefont {R.~K.}\ \bibnamefont
  {Follett}}, \bibinfo {author} {\bibfnamefont {J.~G.}\ \bibnamefont {Shaw}},
  \bibinfo {author} {\bibfnamefont {J.~F.}\ \bibnamefont {Myatt}}, \bibinfo
  {author} {\bibfnamefont {C.}~\bibnamefont {Dorrer}}, \bibinfo {author}
  {\bibfnamefont {D.~H.}\ \bibnamefont {Froula}}, \ and\ \bibinfo {author}
  {\bibfnamefont {J.~P.}\ \bibnamefont {Palastro}},\ }\bibfield  {title}
  {\enquote {\bibinfo {title} {Thresholds of absolute instabilities driven by a
  broadband laser},}\ }\href {\doibase 10.1063/1.5098479} {\bibfield  {journal}
  {\bibinfo  {journal} {Physics of Plasmas}\ }\textbf {\bibinfo {volume}
  {26}},\ \bibinfo {pages} {062111} (\bibinfo {year} {2019})}\BibitemShut
  {NoStop}%
\bibitem [{\citenamefont {Zhao}\ \emph {et~al.}(2021)\citenamefont {Zhao},
  \citenamefont {Wu}, \citenamefont {Weng}, \citenamefont {Sheng},\ and\
  \citenamefont {Zhu}}]{zhaoMitigationMultibeamStimulated2021}%
  \BibitemOpen
  \bibfield  {author} {\bibinfo {author} {\bibfnamefont {Y.}~\bibnamefont
  {Zhao}}, \bibinfo {author} {\bibfnamefont {C.~F.}\ \bibnamefont {Wu}},
  \bibinfo {author} {\bibfnamefont {S.}~\bibnamefont {Weng}}, \bibinfo {author}
  {\bibfnamefont {Z.}~\bibnamefont {Sheng}}, \ and\ \bibinfo {author}
  {\bibfnamefont {J.}~\bibnamefont {Zhu}},\ }\bibfield  {title} {\enquote
  {\bibinfo {title} {Mitigation of multibeam stimulated {{Raman}} scattering
  with polychromatic light},}\ }\href {\doibase 10.1088/1361-6587/abe75a}
  {\bibfield  {journal} {\bibinfo  {journal} {Plasma Physics and Controlled
  Fusion}\ }\textbf {\bibinfo {volume} {63}},\ \bibinfo {pages} {055006}
  (\bibinfo {year} {2021})}\BibitemShut {NoStop}%
\bibitem [{\citenamefont {Ma}\ \emph {et~al.}(2021)\citenamefont {Ma},
  \citenamefont {Li}, \citenamefont {Weng}, \citenamefont {Yew}, \citenamefont
  {Kawata}, \citenamefont {Gibbon}, \citenamefont {Sheng},\ and\ \citenamefont
  {Zhang}}]{weng}%
  \BibitemOpen
  \bibfield  {author} {\bibinfo {author} {\bibfnamefont {H.~H.}\ \bibnamefont
  {Ma}}, \bibinfo {author} {\bibfnamefont {X.~F.}\ \bibnamefont {Li}}, \bibinfo
  {author} {\bibfnamefont {S.~M.}\ \bibnamefont {Weng}}, \bibinfo {author}
  {\bibfnamefont {S.~H.}\ \bibnamefont {Yew}}, \bibinfo {author} {\bibfnamefont
  {S.}~\bibnamefont {Kawata}}, \bibinfo {author} {\bibfnamefont
  {P.}~\bibnamefont {Gibbon}}, \bibinfo {author} {\bibfnamefont {Z.~M.}\
  \bibnamefont {Sheng}}, \ and\ \bibinfo {author} {\bibfnamefont
  {J.}~\bibnamefont {Zhang}},\ }\bibfield  {title} {\enquote {\bibinfo {title}
  {{Mitigating parametric instabilities in plasmas by sunlight-like lasers}},}\
  }\href {\doibase 10.1063/5.0054653} {\bibfield  {journal} {\bibinfo
  {journal} {Matter and Radiation at Extremes}\ }\textbf {\bibinfo {volume}
  {6}} (\bibinfo {year} {2021}),\ 10.1063/5.0054653},\ \bibinfo {note}
  {055902}\BibitemShut {NoStop}%
\bibitem [{\citenamefont {Munro}\ \emph {et~al.}(2004)\citenamefont {Munro},
  \citenamefont {Dixit}, \citenamefont {Langdon},\ and\ \citenamefont
  {Murray}}]{PS1}%
  \BibitemOpen
  \bibfield  {author} {\bibinfo {author} {\bibfnamefont {D.~H.}\ \bibnamefont
  {Munro}}, \bibinfo {author} {\bibfnamefont {S.~N.}\ \bibnamefont {Dixit}},
  \bibinfo {author} {\bibfnamefont {A.~B.}\ \bibnamefont {Langdon}}, \ and\
  \bibinfo {author} {\bibfnamefont {J.~R.}\ \bibnamefont {Murray}},\ }\bibfield
   {title} {\enquote {\bibinfo {title} {Polarization smoothing in a convergent
  beam},}\ }\href {\doibase 10.1364/AO.43.006639} {\bibfield  {journal}
  {\bibinfo  {journal} {Appl. Opt.}\ }\textbf {\bibinfo {volume} {43}},\
  \bibinfo {pages} {6639--6647} (\bibinfo {year} {2004})}\BibitemShut {NoStop}%
\bibitem [{\citenamefont {Skupsky}\ \emph {et~al.}(1989)\citenamefont
  {Skupsky}, \citenamefont {Short}, \citenamefont {Kessler}, \citenamefont
  {Craxton}, \citenamefont {Letzring},\ and\ \citenamefont {Soures}}]{ssd}%
  \BibitemOpen
  \bibfield  {author} {\bibinfo {author} {\bibfnamefont {S.}~\bibnamefont
  {Skupsky}}, \bibinfo {author} {\bibfnamefont {R.~W.}\ \bibnamefont {Short}},
  \bibinfo {author} {\bibfnamefont {T.}~\bibnamefont {Kessler}}, \bibinfo
  {author} {\bibfnamefont {R.~S.}\ \bibnamefont {Craxton}}, \bibinfo {author}
  {\bibfnamefont {S.}~\bibnamefont {Letzring}}, \ and\ \bibinfo {author}
  {\bibfnamefont {J.~M.}\ \bibnamefont {Soures}},\ }\bibfield  {title}
  {\enquote {\bibinfo {title} {{Improved laser‐beam uniformity using the
  angular dispersion of frequency‐modulated light}},}\ }\href {\doibase
  10.1063/1.344101} {\bibfield  {journal} {\bibinfo  {journal} {Journal of
  Applied Physics}\ }\textbf {\bibinfo {volume} {66}},\ \bibinfo {pages}
  {3456--3462} (\bibinfo {year} {1989})},\ \Eprint
  {http://arxiv.org/abs/https://pubs.aip.org/aip/jap/article-pdf/66/8/3456/8009365/3456\_1\_online.pdf}
  {https://pubs.aip.org/aip/jap/article-pdf/66/8/3456/8009365/3456\_1\_online.pdf}
  \BibitemShut {NoStop}%
\bibitem [{\citenamefont {Barth}\ and\ \citenamefont {Fisch}(2016)}]{Ido}%
  \BibitemOpen
  \bibfield  {author} {\bibinfo {author} {\bibfnamefont {I.}~\bibnamefont
  {Barth}}\ and\ \bibinfo {author} {\bibfnamefont {N.~J.}\ \bibnamefont
  {Fisch}},\ }\bibfield  {title} {\enquote {\bibinfo {title} {{Reducing
  parametric backscattering by polarization rotation}},}\ }\href {\doibase
  10.1063/1.4964291} {\bibfield  {journal} {\bibinfo  {journal} {Physics of
  Plasmas}\ }\textbf {\bibinfo {volume} {23}} (\bibinfo {year} {2016}),\
  10.1063/1.4964291},\ \bibinfo {note} {102106}\BibitemShut {NoStop}%
\bibitem [{\citenamefont {Arber}\ \emph {et~al.}(2015)\citenamefont {Arber},
  \citenamefont {Bennett}, \citenamefont {Brady}, \citenamefont
  {{Lawrence-Douglas}}, \citenamefont {Ramsay}, \citenamefont {Sircombe},
  \citenamefont {Gillies}, \citenamefont {Evans}, \citenamefont {Schmitz},
  \citenamefont {Bell},\ and\ \citenamefont {Ridgers}}]{Arber_2015}%
  \BibitemOpen
  \bibfield  {author} {\bibinfo {author} {\bibfnamefont {T.~D.}\ \bibnamefont
  {Arber}}, \bibinfo {author} {\bibfnamefont {K.}~\bibnamefont {Bennett}},
  \bibinfo {author} {\bibfnamefont {C.~S.}\ \bibnamefont {Brady}}, \bibinfo
  {author} {\bibfnamefont {A.}~\bibnamefont {{Lawrence-Douglas}}}, \bibinfo
  {author} {\bibfnamefont {M.~G.}\ \bibnamefont {Ramsay}}, \bibinfo {author}
  {\bibfnamefont {N.~J.}\ \bibnamefont {Sircombe}}, \bibinfo {author}
  {\bibfnamefont {P.}~\bibnamefont {Gillies}}, \bibinfo {author} {\bibfnamefont
  {R.~G.}\ \bibnamefont {Evans}}, \bibinfo {author} {\bibfnamefont
  {H.}~\bibnamefont {Schmitz}}, \bibinfo {author} {\bibfnamefont {A.~R.}\
  \bibnamefont {Bell}}, \ and\ \bibinfo {author} {\bibfnamefont {C.~P.}\
  \bibnamefont {Ridgers}},\ }\bibfield  {title} {\enquote {\bibinfo {title}
  {Contemporary particle-in-cell approach to laser-plasma modelling},}\ }\href
  {\doibase 10.1088/0741-3335/57/11/113001} {\bibfield  {journal} {\bibinfo
  {journal} {Plasma Physics and Controlled Fusion}\ }\textbf {\bibinfo {volume}
  {57}},\ \bibinfo {pages} {113001} (\bibinfo {year} {2015})}\BibitemShut
  {NoStop}%
\bibitem [{\citenamefont {Bauer}\ \emph {et~al.}(2015)\citenamefont {Bauer},
  \citenamefont {Banzer}, \citenamefont {Karimi}, \citenamefont {Orlov},
  \citenamefont {Rubano}, \citenamefont {Marrucci}, \citenamefont {Santamato},
  \citenamefont {Boyd},\ and\ \citenamefont {Leuchs}}]{VLconstruction}%
  \BibitemOpen
  \bibfield  {author} {\bibinfo {author} {\bibfnamefont {T.}~\bibnamefont
  {Bauer}}, \bibinfo {author} {\bibfnamefont {P.}~\bibnamefont {Banzer}},
  \bibinfo {author} {\bibfnamefont {E.}~\bibnamefont {Karimi}}, \bibinfo
  {author} {\bibfnamefont {S.}~\bibnamefont {Orlov}}, \bibinfo {author}
  {\bibfnamefont {A.}~\bibnamefont {Rubano}}, \bibinfo {author} {\bibfnamefont
  {L.}~\bibnamefont {Marrucci}}, \bibinfo {author} {\bibfnamefont
  {E.}~\bibnamefont {Santamato}}, \bibinfo {author} {\bibfnamefont {R.~W.}\
  \bibnamefont {Boyd}}, \ and\ \bibinfo {author} {\bibfnamefont
  {G.}~\bibnamefont {Leuchs}},\ }\bibfield  {title} {\enquote {\bibinfo {title}
  {Observation of optical polarization m\"bius strips},}\ }\href {\doibase
  10.1126/science.1260635} {\bibfield  {journal} {\bibinfo  {journal}
  {Science}\ }\textbf {\bibinfo {volume} {347}},\ \bibinfo {pages} {964--966}
  (\bibinfo {year} {2015})},\ \Eprint
  {http://arxiv.org/abs/https://www.science.org/doi/pdf/10.1126/science.1260635}
  {https://www.science.org/doi/pdf/10.1126/science.1260635} \BibitemShut
  {NoStop}%
\bibitem [{\citenamefont {Padgett}(2017)}]{Padgett2017OrbitalAM}%
  \BibitemOpen
  \bibfield  {author} {\bibinfo {author} {\bibfnamefont {M.~J.}\ \bibnamefont
  {Padgett}},\ }\bibfield  {title} {\enquote {\bibinfo {title} {Orbital angular
  momentum 25 years on [invited].}}\ }\href@noop {} {\bibfield  {journal}
  {\bibinfo  {journal} {Optics express}\ }\textbf {\bibinfo {volume} {25 10}},\
  \bibinfo {pages} {11265--11274} (\bibinfo {year} {2017})}\BibitemShut
  {NoStop}%
\bibitem [{\citenamefont {Wen}\ \emph {et~al.}(2019)\citenamefont {Wen},
  \citenamefont {Maximov}, \citenamefont {Yan}, \citenamefont {Li},
  \citenamefont {Ren},\ and\ \citenamefont
  {Tsung}}]{wenThreedimensionalParticleincellModeling2019}%
  \BibitemOpen
  \bibfield  {author} {\bibinfo {author} {\bibfnamefont {H.}~\bibnamefont
  {Wen}}, \bibinfo {author} {\bibfnamefont {A.~V.}\ \bibnamefont {Maximov}},
  \bibinfo {author} {\bibfnamefont {R.}~\bibnamefont {Yan}}, \bibinfo {author}
  {\bibfnamefont {J.}~\bibnamefont {Li}}, \bibinfo {author} {\bibfnamefont
  {C.}~\bibnamefont {Ren}}, \ and\ \bibinfo {author} {\bibfnamefont {F.~S.}\
  \bibnamefont {Tsung}},\ }\bibfield  {title} {\enquote {\bibinfo {title}
  {Three-dimensional particle-in-cell modeling of parametric instabilities near
  the quarter-critical density in plasmas},}\ }\href {\doibase
  10.1103/PhysRevE.100.041201} {\bibfield  {journal} {\bibinfo  {journal}
  {Physical Review E}\ }\textbf {\bibinfo {volume} {100}},\ \bibinfo {pages}
  {041201} (\bibinfo {year} {2019})}\BibitemShut {NoStop}%
\bibitem [{\citenamefont {Xiao}\ \emph {et~al.}(2020)\citenamefont {Xiao},
  \citenamefont {Zhuo}, \citenamefont {Yin}, \citenamefont {Liu}, \citenamefont
  {Zheng},\ and\ \citenamefont {He}}]{xiaoTransitionTwoplasmonDecay2020}%
  \BibitemOpen
  \bibfield  {author} {\bibinfo {author} {\bibfnamefont {C.}~\bibnamefont
  {Xiao}}, \bibinfo {author} {\bibfnamefont {H.}~\bibnamefont {Zhuo}}, \bibinfo
  {author} {\bibfnamefont {Y.}~\bibnamefont {Yin}}, \bibinfo {author}
  {\bibfnamefont {Z.}~\bibnamefont {Liu}}, \bibinfo {author} {\bibfnamefont
  {C.}~\bibnamefont {Zheng}}, \ and\ \bibinfo {author} {\bibfnamefont
  {X.}~\bibnamefont {He}},\ }\bibfield  {title} {\enquote {\bibinfo {title}
  {Transition from two-plasmon decay to stimulated {{Raman}} scattering under
  ignition conditions},}\ }\href {\doibase 10.1088/1741-4326/ab4e79} {\bibfield
   {journal} {\bibinfo  {journal} {Nuclear Fusion}\ }\textbf {\bibinfo {volume}
  {60}},\ \bibinfo {pages} {016022} (\bibinfo {year} {2020})}\BibitemShut
  {NoStop}%
\bibitem [{\citenamefont {Pan}\ \emph {et~al.}(2018)\citenamefont {Pan},
  \citenamefont {Jiang}, \citenamefont {Wang}, \citenamefont {Guo},
  \citenamefont {Li}, \citenamefont {Li}, \citenamefont {Yang}, \citenamefont
  {Zheng}, \citenamefont {Zhang},\ and\ \citenamefont
  {He}}]{panTwoplasmonDecayInstability2018}%
  \BibitemOpen
  \bibfield  {author} {\bibinfo {author} {\bibfnamefont {K.~Q.}\ \bibnamefont
  {Pan}}, \bibinfo {author} {\bibfnamefont {S.~E.}\ \bibnamefont {Jiang}},
  \bibinfo {author} {\bibfnamefont {Q.}~\bibnamefont {Wang}}, \bibinfo {author}
  {\bibfnamefont {L.}~\bibnamefont {Guo}}, \bibinfo {author} {\bibfnamefont
  {S.~W.}\ \bibnamefont {Li}}, \bibinfo {author} {\bibfnamefont {Z.~C.}\
  \bibnamefont {Li}}, \bibinfo {author} {\bibfnamefont {D.}~\bibnamefont
  {Yang}}, \bibinfo {author} {\bibfnamefont {C.~Y.}\ \bibnamefont {Zheng}},
  \bibinfo {author} {\bibfnamefont {B.~H.}\ \bibnamefont {Zhang}}, \ and\
  \bibinfo {author} {\bibfnamefont {X.~T.}\ \bibnamefont {He}},\ }\bibfield
  {title} {\enquote {\bibinfo {title} {Two-plasmon decay instability of the
  backscattered light of stimulated {{Raman}} scattering},}\ }\href {\doibase
  10.1088/1741-4326/aad059} {\bibfield  {journal} {\bibinfo  {journal} {Nuclear
  Fusion}\ }\textbf {\bibinfo {volume} {58}},\ \bibinfo {pages} {096035}
  (\bibinfo {year} {2018})}\BibitemShut {NoStop}%
\bibitem [{\citenamefont {Yan}, \citenamefont {Maximov},\ and\ \citenamefont
  {Ren}(2010)}]{RuiYanTPDPOP}%
  \BibitemOpen
  \bibfield  {author} {\bibinfo {author} {\bibfnamefont {R.}~\bibnamefont
  {Yan}}, \bibinfo {author} {\bibfnamefont {A.~V.}\ \bibnamefont {Maximov}}, \
  and\ \bibinfo {author} {\bibfnamefont {C.}~\bibnamefont {Ren}},\ }\bibfield
  {title} {\enquote {\bibinfo {title} {{The linear regime of the two-plasmon
  decay instability in inhomogeneous plasmas}},}\ }\href {\doibase
  10.1063/1.3414350} {\bibfield  {journal} {\bibinfo  {journal} {Physics of
  Plasmas}\ }\textbf {\bibinfo {volume} {17}} (\bibinfo {year} {2010}),\
  10.1063/1.3414350},\ \bibinfo {note} {052701}\BibitemShut {NoStop}%
\bibitem [{\citenamefont {Afeyan}\ and\ \citenamefont
  {Williams}(1985)}]{afeyanStimulatedRamanSidescattering1985}%
  \BibitemOpen
  \bibfield  {author} {\bibinfo {author} {\bibfnamefont {B.~B.}\ \bibnamefont
  {Afeyan}}\ and\ \bibinfo {author} {\bibfnamefont {E.~A.}\ \bibnamefont
  {Williams}},\ }\bibfield  {title} {\enquote {\bibinfo {title} {Stimulated
  {{Raman}} sidescattering with the effects of oblique incidence},}\ }\href
  {\doibase 10.1063/1.865340} {\bibfield  {journal} {\bibinfo  {journal}
  {Physics of Fluids}\ }\textbf {\bibinfo {volume} {28}},\ \bibinfo {pages}
  {3397} (\bibinfo {year} {1985})}\BibitemShut {NoStop}%
\end{thebibliography}%

\end{document}